\documentclass[a4paper,aps,prx,twocolumn,floatfix,superscriptaddress,notitlepage,usenames,dvipsnames,svgnames,table,footinbib,longbibliography]{quantumarticle}
\pdfoutput=1
\usepackage[utf8]{inputenc}
\usepackage[T1]{fontenc}
\usepackage{graphicx}
\usepackage{grffile}
\usepackage{wrapfig}
\usepackage{rotating}
\usepackage[normalem]{ulem}
\usepackage{amsmath}
\usepackage{textcomp}
\usepackage{amssymb}
\usepackage{capt-of}

\usepackage{parskip}
\usepackage{mathrsfs}
\usepackage[margin= 0.75in]{geometry}
\usepackage[braket, qm]{qcircuit}

\usepackage[english]{babel}
\usepackage{letltxmacro}
\LetLtxMacro{\ORIGselectlanguage}{\selectlanguage}
\makeatletter
\DeclareRobustCommand{\selectlanguage}[1]{%
  \@ifundefined{alias@\string#1}
    {\ORIGselectlanguage{#1}}
    {\begingroup\edef\x{\endgroup
      \noexpand\ORIGselectlanguage{\@nameuse{alias@#1}}}\x}%
}
\newcommand{\definelanguagealias}[2]{%
  \@namedef{alias@#1}{#2}%
}
\makeatother
\definelanguagealias{en}{english}
\definelanguagealias{eng}{english}

\usepackage{bbm}
\usepackage{etoolbox}
\usepackage{tcolorbox} 
\usepackage{tikz}
\usepackage{amsthm}
\usepackage{amssymb}
\usetikzlibrary{arrows.meta}
\usepackage{mathtools}

\usepackage{bm}
\usepackage{dsfont}
\usepackage[font=small,labelfont=bf,justification=RaggedRight,format=plain]{caption}
\usepackage{subcaption}

\usepackage{thmtools}
\usepackage{thm-restate}

\usepackage{microtype}
\usepackage{dsfont}
\usepackage{tensor}

\usepackage{physics}
\usepackage{mathrsfs}
\usepackage{mathtools}
\usepackage{fixltx2e}
\usepackage{relsize}

\usepackage{float}
\usepackage[Export]{adjustbox} 
\adjustboxset{max size={0.9\linewidth}{0.9\paperheight}}

\usepackage[font=small,labelfont=bf,justification=RaggedRight,format=plain]{caption}
\usepackage{subcaption}

\DeclarePairedDelimiterX\phys[2]{\langle}{\rangle}{#1 \delimsize\vert\mathopen{} #2}

\theoremstyle{remark}

\definecolor{blue-violet}{rgb}{0.54, 0.17, 0.89}
\usepackage{hyperref}
\hypersetup{
 pdfauthor={Paolo Zanardi et al},
 pdftitle={Operational Quantum Mereology and Minimal Scrambling}
 pdflang={English},colorlinks=true,linkcolor=RubineRed,citecolor=blue-violet,urlcolor=Cerulean} 
\usepackage[capitalise]{cleveref}

\begin{document}

\title{Operational Quantum Mereology and Minimal Scrambling}

\author{Paolo Zanardi}
\email [e-mail: ]{zanardi@usc.edu}
\affiliation{Department of Physics and Astronomy, and Center for Quantum Information Science and Technology, University of Southern California, Los Angeles, California 90089-0484, USA}
\affiliation{Department of Mathematics, University of Southern California, Los Angeles, California 90089-2532, USA}
\author{Emanuel Dallas}
\email [e-mail: ]{dallas@usc.edu}
\affiliation{Department of Physics and Astronomy, and Center for Quantum Information Science and Technology, University of Southern California, Los Angeles, California 90089-0484, USA}
\author{Faidon Andreadakis}
\email [e-mail: ]{fandread@usc.edu}
\affiliation{Department of Physics and Astronomy, and Center for Quantum Information Science and Technology, University of Southern California, Los Angeles, California 90089-0484, USA}
\author{Seth Lloyd}
\email [e-mail: ]{slloyd@mit.edu}
\affiliation{MIT Department of Mechanical Engineering, 77 Massachusetts Avenue, Cambridge, MA 02139, USA}
\affiliation{Turing Inc., Cambridge, MA 02139, USA}

\date{July 7, 2024}
\begin{abstract}
In this paper we will attempt to answer the following question: what are the natural quantum subsystems which emerge out of a system’s dynamical laws?  To answer this question we first define generalized tensor product structures (gTPS) in terms of observables,  as dual pairs of an operator subalgebra $\cal A$ and its commutant.
Second, we propose an operational  criterion of {{minimal information scrambling}} at short time scales to dynamically  select gTPS.  In this way the emergent subsystems are those which maintain their ``informational identities'' the longest.  
This strategy is made quantitative by defining a Gaussian scrambling rate in terms of the short-time expansion of an algebraic version of the Out of Time Order Correlation (OTOC) function i.e., the $\cal A$-OTOC.  The Gaussian scrambling rate is computed analytically for physically important cases of general division into subsystems and is shown to have an intuitive and compelling physical interpretation in terms of minimizing the interaction strength between subsystems.
\end{abstract}

\maketitle
\section{Introduction}\label{sec:intro}
%
Mereology is the theory of parthood relations: the relations of part to whole and the relations of part to part within a whole.  In this paper we shall try taking some steps towards an information-theoretic and operational framework for {\em{quantum mereology} }\cite{carroll-mereology-2021}.

It is part and parcel of the reductionistic approach of modern science to explain the behavior of complex systems in terms of their simpler constituents, e.g. particles,  their properties and their mutual interactions.  However,  such a division into simpler components is by {\em{no means}} unique as it depends on the questions one is trying to address, the experimental limitations of the observer, and the physical regime one is operating in. {Some related elementary paradigms exist in various fields of theoretical physics; for example, what comprises the elementary parts in scattering experiments or the decoupled (``frozen'') degrees of freedom in statistical mechanics depends on the energy scale under consideration.}

This freedom elicits the following general questions: how does one select, among manifold possibilities, the relevant subdivision of a system into sub-parts? Can one establish a compelling connection between the intrinsic dynamics of the system, the operational capabilities of the observer, and a ``naturally'' emergent multi-partite structure? This paper makes an attempt at answering these  sweeping questions by {employing} ideas from quantum information and using some elementary operator algebra tools.  
Our work takes as its starting point the pathbreaking papers of Zurek  \cite{zurek-sieve}, and Carroll and Singh  \cite{carroll-mereology-2021}. Zurek introduced the so-called ``predictability sieve'' \cite{zurek-sieve} procedure for the emergence of classicality in systems coupled with an environment. In \cite{zurek-sieve}, Zurek showed how semi-classical pointer variables emerge from the interaction between a system and its environment: the pointer variables are those whose time evolution is most predictable given the system dynamics. In that work, the division into system and environment was assumed to be given. In \cite{carroll-mereology-2021}, Carroll and Singh extended Zurek's criterion to identify the division into system and environment that gives the maximally predictable pointer variables.   
These works emphasize predictability as a criterion for identifying good bipartitions. Our work charts a novel course that is inspired by this approach, but that extends it in significant ways. First, our framework is purely quantum, unconcerned with emergent classicality. Second, we consider more generalized subdivisions beyond system-environment bipartitions. Our mereological approach depends solely on the system dynamics -- given operational constraints, it gives the partition \textit{selected} by the Hamiltonian. In this sense, the emergent partitions that our approach unveils are fundamental from a dynamical point of view. Finally, by emphasizing  {\it scrambling} as a measure of unpredictability, we can bring to play powerful methods from the extensive recent literature on scrambling theory. We also note the existence of Hamiltonian/spectral approaches for identifying bi- and multi-partitions \cite{Cotler2019-qq}.
Let us start by motivating our particular strategy with a qualitative analogy. {Consider the system of a soluble, macroscopic solid inserted into a liquid solvent, inside a glass beaker. One can, of course, subdivide this system into the ``background'' partition of individual molecules. However, assuming one can only observe this system without advanced equipment, this subdivision is useless. Individual molecules change location and relative orientation, and exchange energy, much too quickly to perceive them at such a granular level. Naturally, one divides this system into the components that can be perceived as relevantly different: the beaker, the solvent, and the solid. Note that this ``tripartition" does not last forever -- eventually, the solid fully dissolves, becomes aqueous, and mixes amongst the solvent. After enough time, the solvent in the beaker evaporates, and after eons, the beaker itself disintegrates. In spite of this, the tripartition minimizes the rate at which its constituent components ``leak'' into one another.}

{We carry this intuition over to general quantum systems in which subsystems of a given partition might not \textit{spatially} mix, but information can scramble between them.} We  will  {therefore} use an operational  criterion of {\em{minimal information scrambling {rate}}} at short time scales to dynamically  select generalized partitions.  This strategy amounts to saying that the emergent subsystems are those {that have their informational content scrambled to ``external'' degrees of freedom the slowest by the dynamics, hence maintaining their internal ``informational identity'' the longest.} We are now in the position of outlining the  approach we will pursue in this paper in a more technical fashion.

The first fundamental  ingredient of  this paper is provided by the algebraic approach to quantum {\em{virtual  quantum subsystems}} (VQS) originally advocated in \cite{zanardi-virtual-2001,zanardi-tps-2004}. 
The fundamental idea is that operational constraints may limit the set of physically relevant observables and operations to a subalgebra $\cal A$ of operators, which in turn induces a decomposition of the state space into a {\em{direct sum}} of virtual quantum subsystems. This generalized tensor product structure (gTPS)  
  plays a key role in the theory of decoherence-free subspaces  \cite{zanardi-noiseless-1997,lidar-dfs-1998},
noiseless subsystems \cite{KLV-2000,zanardi-stabilizing-2000}, quantum reference frames \cite{QRF-RMP-2007}, topological protection \cite{NS-topo}, and operator error correction \cite{OEC-Kribs-2004}. %
 Some interesting  recent developments of the general theory can be found in \cite{kabernik-coarse-2018,kabernik-reduction-2020}.

The second fundamental ingredient is the algebraic out-of-time-order commutator, ``$\cal A$-OTOC,"  approach to {\em{quantum information scrambling}} introduced in \cite{Zanardi-GAAC-2021, A-OTOC-Faidon-2022}.
The $\cal A$-OTOC {quantifies the scrambling of information stored in the physical degrees of freedom represented by a subalgebra of observables $\mathcal{A}$, and it therefore serves as a natural quantity from which we can derive a scrambling rate. It has previously proved successful in furthering the understanding of} two seemingly unrelated physical problems -- operator entanglement \cite{zanardi2001entanglement} 
\cite{yan_information_2019,styliaris_information_2020} and coherence-generating power (CGP) \cite{zanardiCoherencegeneratingPowerQuantum2017, zanardi-CGP-measures-2007} -- from a single theoretical vantage point.

It is of paramount importance to stress that  such an algebra-based  strategy  can be applied to a variety  of situations in which there is not an \emph{a priori} locality structure which  gives a natural way of defining subsystems, e.g. \cite{carroll-mereology-2021} in the context of quantum gravity, and, through operator error correction, holography see \cite{Harlow,Akers}. {In cases \textit{with} ``background'' locality structures, the emergent system partition may substantially differ from the background -- the final example in \ref{sec:rate}} provides an illustration of this. Also, {we note} our approach is very close in spirit to the algebraic approach to quantum field theory (QFT) -- see, e.g., Haag's classic book \cite{haagLocalQuantumPhysics1996} and the recent \cite{Witten-Entanglement} on entanglement in QFT. However, in this paper we will be focusing on finite-dimensional Hilbert spaces which are more relevant to most of quantum information.

The paper is structured as follows. Sect (\ref{sec:preliminaries}) contains some introductory material on notation and  algebras. 
Sect. (\ref{sec:generalized}) discusses the gTPS concept used in this paper. 
Sect. (\ref{sec:A-OTOC}) 
reviews definitions and basic results on the $\cal A$-OTOC formalism for scrambling of algebras. In this section, we introduce three physical examples, which we subsequently follow throughout the paper: the first example is the scrambling rate of a subsystem; the second is the coherence generating power of quantum dynamics; the third is the example of quantum error correcting codes, which by definition suppress scrambling within the code space.   In Sect (\ref{sec:rate}) the notion of Gaussian scrambling rate is introduced and deployed to define dynamical emergence of gTPS for these examples.   In the first example, we provide a proof for the intuitive result that the division into subsystems that minimizes subsystem scrambling is the division that minimizes the norm of the interaction Hamiltonian. For case 2, we bound the Gaussian scrambling rate by the coherence generating power of maximal commuting subalgebras.   For case 3, we relate scrambling rate to the detection and correction of errors in a stabilizer code.   Finally, Sect. (\ref{sec:conclusions}) contains conclusions and outlook. The mathematical proofs of most of the results are in the Appendix.

\section{Preliminaries}\label{sec:preliminaries}
Let ${\cal H}$
be a $d$-dimensional Hilbert space and $L({\cal H}) $ its full operator algebra. 
$L({\cal H}) $  has  a Hilbert space structure via  the Hilbert-Schmidt scalar product: $\langle X,\,Y\rangle:=\mathrm{Tr}\left(X^\dagger Y\right)$ and norm
$\|X\|_2^2:=\langle X,\, X\rangle.$
This equips the space of superoperators i.e., $L(L({\cal H}))$ with the scalar product $\langle {\cal T},\,{\cal F}\rangle:=\mathrm{Tr}_{HS}\left({\cal T}^\dagger {\cal F} \right)=
\sum_{l,m} \langle m| {\cal T}^\dagger {\cal F}(|m\rangle\langle l|)|l\rangle,$ and the norm $$\|  {\cal T}\|_{HS}^2= \langle {\cal T},\,{\cal T}\rangle=\sum_{l,m} \|{\cal T}(|m\rangle\langle l|)\|_2^2.$$
For example, if ${\cal T}(X)=\sum_i A_i XA_i^\dagger,$ then $\|{\cal T}\|_{HS}^2=\sum_{i,j} |\mathrm{Tr}(A_i^\dagger A_j)|^2.$
Moreover, if $\mathbb{P}$ is an orthogonal  projection, its rank is given by  $\|\mathbb{P}\|_{HS}^2=\mathrm{Tr}_{HS} \mathbb{P}.$

The key  objects of this paper are   \emph{hermitian-closed  unital  subalgebras}\footnote{{Namely, they contain the identity and it holds that $X\in \mathcal{A} \rightarrow X^\dagger \in \mathcal{A}$}} ${\cal A}\subset L({\cal H})$ and their {\em{commutants}}
\begin{align}\label{eq:commutant}
{\cal A}^\prime:=\{X\in L({\cal H})\,|\, [X,\,Y]=0,\,\forall\, Y\in {\cal A}\}
\end{align}
and {\em{centers}} 
${\cal Z}({\cal A}):={\cal A}\cap {\cal A}^\prime.
$

For these algebras, the {\em{double commutant}} theorem holds: $({\cal A}^\prime)^\prime={\cal A}.$
If $\cal A$ (${\cal A}^\prime$) is abelian, one has that ${\cal A}\subset {\cal A}^\prime$ (${\cal A}^\prime\subset {\cal A}$)
and therefore ${\cal Z}({\cal A})={\cal A}$ (${\cal Z}({\cal A})={\cal A}^\prime$).

The fundamental structure theorem of these objects states that the Hilbert
space breaks into a direct sum of $d_Z:=\mathrm{dim} \,{\cal Z}({\cal A})$ blocks  and each of them has a tensor product bipartite structure 
\begin{align}\label{eq:hilb-decomp}
{\cal H}=\bigoplus_{J=1}^{d_Z}  {\cal H}_J,\qquad {\cal H}_J\cong \mathbf{C}^{n_J}\otimes  \mathbf{C}^{d_J}.
\end{align}
The factors  $\mathbf{C}^{n_J}$ and $\mathbf{C}^{d_J}$ are  VQS \cite{zanardi-virtual-2001}.
Moreover, at the algebra level one has that $\cal A$ (${\cal A}^\prime$) acts irreducibly on the $  \mathbf{C}^{d_J}$ factors ($\mathbf{C}^{n_J}$)
${\cal A}\cong \bigoplus_{J=1}^{d_Z}  \mathbf{1}_{n_J}\otimes L( \mathbf{C}^{d_J}),\quad{\cal A}^\prime\cong \bigoplus_{J=1}^{d_Z}  L( \mathbf{C}^{n_J})\otimes \mathbf{1}_{d_J}.
$
 Whence, $d=\sum_J n_Jd_J,$ and
 $$\mathrm{dim}\,{\cal A}=\sum_Jd_J^2=:d({\cal A}),\quad\mathrm{dim}\,{\cal A}^\prime=\sum_Jn_J^2=:d({\cal A}^\prime).\nonumber
 $$
If $\cal A$ (${\cal A}^\prime$) is {{abelian}} then $d_J=1,\,(\forall J)$  ($n_J=1,\,(\forall J)$).
By defining the $d_Z$-dimensional (integer-valued) vectors ${\bf{d}}:=(d_J)_J$, and ${\bf{n}}:=(n_J)_J$, one sees that
 $$d^2=(\mathbf{n}\cdot\mathbf{d})^2\le \|\mathbf{n}\|^2\|\mathbf{d}\|^2=
 d({\cal A})d({\cal A}^\prime).$$ 
If  ${\bf{d}}=\lambda {\bf{n}}$, the above inequality becomes an equality, i.e. $d^2=d({\cal A})d({\cal A}^\prime),$
 In this case the pair $({\cal A},\,{\cal A}^\prime)$ is referred to as  \emph{collinear}.
The center of $\cal A$ is spanned by the projections over the central blocks
 $${\cal Z}({\cal A})=\bigoplus_{J=1}^{d_Z}  \mathbf{C}\{\Pi_J:=\mathbf{1}_{n_J}\otimes \mathbf{1}_{d_J}\}.$$
%
%
%
Associated to any algebra $\cal A$ is a completely positive (CP) orthogonal projection  map: $\mathbb{P}_{\cal A}^\dagger=\mathbb{P}_{\cal A},\, \mathbb{P}_{\cal A}^2=\mathbb{P}_{\cal A}$ and
$\mathrm{Im}\,\mathbb{P}_{\cal A}={\cal A}$ \footnote{
In terms of (\ref{eq:hilb-decomp}) one has 
$\mathbb{P}_{{\cal A}}(X) =\sum^\oplus_J \frac{\mathbf{1}_{n_J}}{n_J} \otimes \mathrm{tr}_{n_J}(X),$ and $\mathbb{P}_{{\cal A}^\prime}(X) =\sum^\oplus_J \mathrm{tr}_{d_J}(X)\otimes \frac{\mathbf{1}_{d_J}}{d_J}.$}.
 Using these  projections, one can define a distance between two algebras $\cal A$ and $\cal B$: 
\begin{align}\label{eq:distance-of-algebras}
 D({\cal A},\,{\cal B}):=\|\mathbb{P}_{\cal A}- \mathbb{P}_{\cal B}\|_{HS}.
\end{align} 
If $U$ is unitary, we denote by ${\cal U}(X):=U XU^\dagger$ and ${\cal U}({\cal A}):=\{ {\cal U}(a)\,|\, a\in{\cal A}\}$ is the image algebra.  The associated projection is given by
 $\mathbb{P}_{{\cal U}({\cal A})}={\cal U} \mathbb{P}_{{\cal A}} {\cal U}^\dagger.$
\section{Generalized TPS}\label{sec:generalized}
{Formally},  the standard  ``tensor product axiom'' of quantum theory says that if we consider the system obtained by the union of two systems, $A$ and $B$, then the  associated Hilbert Space ${\cal H}_{}$ is given by the tensor product of the Hilbert spaces associated with the individual subsystems,
i.e., ${\cal H}_{}\cong {\cal H}_{A}\otimes {\cal H}_{B}.$ At the level of observables, one postulates $L({\cal H}_{})\cong L({\cal H}_{A})\otimes L({\cal H}_{B})$; namely, the full operator algebra is the tensor product of the subsystems operator algebras.
By denoting ${\cal A}:=L({\cal H}_{A})\otimes{\mathbf{1}}_B,$ and ${\cal A}^\prime:={\mathbf{1}}_A\otimes L({\cal H}_{B})$,  one can {equivalently} write
\begin{align}\label{eq:TPS-0}
 {\cal A}\otimes {\cal A}^\prime \cong {\cal A}\vee {\cal A}^\prime\cong L({\cal H}_{}),
\end{align}
where the ${\cal A}\vee {\cal A}^\prime$ denotes the algebra generated by the subalgebras $\cal A$ {\em{and}}  ${\cal A}^\prime$, i.e.,  ${\cal A}\vee {\cal A}^\prime:=\{\sum_i a_i b_i\,|\, a_i\in{\cal A} ,\, b_i\in{\cal A}^\prime\}.$
Now, if  ${\cal A}$ is {\em{any}}  hermitian-closed, unital  subalgebra  of $L({\cal H})$ and  (\ref{eq:TPS-0}) is fulfilled, we say that the dual pair $({\cal A},\, {\cal A}^\prime)$  defines a {\em{virtual bipartition}} or a Tensor Product Structure (TPS) \cite{zanardi-virtual-2001}.

The latter are by no means unique {since} every unitary mapping $U$ over $\cal H$ gives rise to a dual pair $({\cal A}_U\, {\cal A}_U^\prime)$ ($ {\cal A}_U:={\cal U}({\cal A})$) which satisfies  (\ref{eq:TPS-0}). Which one of these TPSs is the ``real" or ``physical one'' depends, again, on the operationally available resources. It is important to emphasize that this is true  even when one knows that ${\cal H}={\cal H}_A\otimes {\cal H}_B$, as the local algebras might  not be  implementable.
If one has control only on $\cal A$, then the associated physical degrees of freedom can be seen as the ``system'' whereas those supported by ${\cal A}^\prime$ can be seen as the ``environment'' (and vice versa). Also, ${\cal A}^\prime$ comprises the ``symmetries'' of $\cal A$ (and vice versa).

In this paper we will consider an even more general situation. Suppose that operational constraints on physical resources   
lead  to the selection of a specific class of realizable operators (observables and operations) forming an algebra $\cal A$
which does {\em{not}} fulfill  (\ref{eq:TPS-0})  but  just 
\begin{align}\label{eq:TPS-1}
{\cal A}\vee {\cal A}^\prime\subset L({\cal H}_{}).
\end{align}
%
 %
 In this case we say that the dual pair $({\cal A},\, {\cal A}^\prime)$  defines a {\em {generalized}} TPS (gTPS).
While (\ref{eq:TPS-1}) is always satisfied, in order to meet the standard condition  (\ref{eq:TPS-0}) $\cal A$ {must be} a {\em{factor}}, i.e., the center 
$ {\cal Z}({\cal A})$ contains just scalars or, equivalently, there is just one $J$ in the decomposition (\ref{eq:hilb-decomp}) {(see \cref{appendix_factor} for further details)}. On the other hand, when $\mathcal{A}^\prime$ is abelian,
then $n_J=1\,(\forall J)$, and the decomposition (\ref{eq:hilb-decomp}) is a form of superselection with the elements of  $ {\cal Z}({\cal A})={\cal A}^\prime$ playing the role of ``superselection charges.''
\par {A formal example of a structure similar to that of Eq.~(\ref{eq:hilb-decomp}) is the Fock space of fermionic particles
\begin{equation}\label{eq:fock}
    F_{-}(\mathcal{H}_1)=\bigoplus_{N =0}^\infty S_{-} \mathcal{H}_1^{\otimes N}\cong \bigotimes_{i=1}^n \mathbb{C}^2
\end{equation}
where $S_{-}$ is the operator that antisymmetrizes the N\textsubscript{th} tensor power $\mathcal{H}_1^{\otimes N}$ of the $n$-dimensional 1-fermion Hilbert space $\mathcal{H}_1$. The isomorphism in Eq. (\ref{eq:fock}) is non-unique; the subsystems $\mathbb{C}^2$ correspond to the fermionic modes \{$c_i, \, i=1,\dots,n\}$ that can be constructed in infinitely many different ways depending on the single-particle basis one chooses to adopt \cite{zanardiQuantumEntanglementFermionic2002}. If the Hamiltonian $H$ of the fermionic system is quadratic, then there are fermionic modes that diagonalize $H$, which arguably provide a ``naturally'' emerging subdivision of the Fock space. More specifically, the Hamiltonian becomes a sum of terms $c_i c_i^\dagger$ that are associated with the different mode subalgebras generated by $c_i,c_i^\dagger, \mathbf{1}$. In this way, each mode gives rise to a dynamically independent subsystem, and information stored in each of the modes does not leak out to other modes.}
\par We would like to stress that we regard  both the tensor {{and}} the additive structure in Eq.~(\ref{eq:hilb-decomp}) as elements of quantum mereology, which in this way conceptually unifies these two aspects.
The first one is the partition into subsystems while the second describes the  partition of the state-space into central symmetry sectors. The latter are labelled by the the super-selection charges' joint eigenvalues and are dynamically decoupled insofar as the operations belong to ${\cal A}$ \footnote{For a CP map this means that the associated Krauss operators are in $\cal A$.}.

\subsection{Noiseless Subsystems for  collective decoherence}\label{subsec:DFS}
To illustrate these concepts, let us consider symmetry-protected quantum information \cite{zanardi-stabilizing-2000}. In this case Eq.~(\ref{eq:TPS-0}) is {\em{not}} satisfied as the relevant algebra is not a factor, and therefore the state-space splits into disjoint symmetry sectors. 
 Consider a system comprising $N$ physical qubits in which the only allowed operations are {\em{global}}, {due to} the lack of spatial resolution. The algebra $\cal  A$  is given by the operators invariant under qubit permutations. It turns out that this is the unital algebra is  generated by the collective spin operators
$S^\alpha:=\sum_{j=1}^N \sigma^\alpha_j,\,(\alpha=x,y,z)$ and that the Hilbert {space} ${\cal H}=(\mathbf{C}^2)^{\otimes N}$ breaks down into $d_J=2J+1$-dimensional  irreducible  components, each with a multiplicity $n_J,$
labelled  by the total angular momentum  $J$: $ {\cal H}=\oplus_J  {\cal H}_J.$  For any value of the quantum number $J$, {there is a} virtual bipartition  ${\cal H}_J\cong \mathbf{C}^{n_J}\otimes  \mathbf{C}^{d_J}.$
In the context  of decoherence protection, $\cal A$ is the ``interaction'' algebra generated by the system's operators coupled to the environment. The $ \mathbf{C}^{n_J}$'s are acted on trivially by $\cal A$ and therefore provide ``virtual noiseless subsystems'' where information can be encoded and safely stored \cite{KLV-2000}. These noiseless subsystems can then be (universally) acted upon by elements of the commutant ${\cal A}^\prime,$ the latter given by the group algebra of permutations generated by pair-wise Heisenberg interactions.
The center ${\cal Z}({\cal A})$ is just the abelian algebra generated by the total angular momentum operator $S^2:=\sum_{\alpha=x,yz} (S^\alpha)^2.$
\section{Scrambling of Algebras.}\label{sec:A-OTOC}
{Let us first illustrate the importance of algebras of observables and commutativity in a simple information-processing setup. Consider two agents, A and B, that have access to the two commuting subalgebras of observables $\mathcal{A}$ and $\mathcal{B}=\mathcal{A}^\prime$ respectively. If the agents have access to a unitary channel $\mathcal{U}$, such that $[X,\mathcal{U}(Y)]=0 \; \forall \,X\in \mathcal{A}, \; \forall \, Y\in \mathcal{B} $ (in words, ``commutativity persists''), then no information can be broadcasted from the physical degrees of freedom associated with $\mathcal{A}$ to
those associated with $\mathcal{B}$ (see \cref{appendix_com}). This shows that studying the commutator between elements of $\mathcal{A}$ and $\mathcal{U}(\mathcal{A}^\prime )$ provides insights into the scrambling of information from $\mathcal{A}$ to $\mathcal{A}^\prime$. }
\par The central object of our algebraic approach to quantum scrambling is the $\mathbf{\cal A}$-OTOC. This tool was {originally} introduced in \cite{Zanardi-GAAC-2021} for closed systems and extended to open ones in \cite{A-OTOC-Faidon-2022}. It is defined as
\begin{align}\label{eq:A-OTOC-0} 
{G}_{{\cal A}}({\cal U}):= \frac{1}{2d}\, \mathbb{E}_{X\in{\cal A},\,Y\in{\cal A}^\prime}\left[\|[X,\,{\cal U}(Y)]\|_2^2\right],
\end{align}
where ${\cal U}(X):=U^\dagger X U$ (with $U$ unitary over $\cal H$).
Here $ \mathbb{E}$ denotes the Haar-average over the unitaries $X$ ($Y$)  in $\cal A$ (${\cal A}^\prime$). 
\textbf{Roughly,} the $\cal A$-OTOC  measures how much the symmetries of the VQS associated to $\cal A$ are dynamically broken by ${\cal U}.$ 
From Eq.~(\ref{eq:A-OTOC-0}) it follows immediately that ${G}_{{\cal A}}({\cal U})={G}_{{\cal A}^\prime}({\cal U}^\dagger)$. Moreover, in the collinear case, one has
${G}_{{\cal A}}={G}_{{\cal A}^\prime}$, and therefore ${G}_{{\cal A}}({\cal U})=G_{{\cal A}}({\cal U}^\dagger)$.

In the collinear case, the $\cal A$-OTOC has a simple geometrical interpretation in terms of the distance between algebras defined  in Eq.~(\ref{eq:distance-of-algebras}) \cite{Zanardi-GAAC-2021}
\begin{align}
{G}_{{\cal A}}({\cal U})=\frac{1}{2} \frac{D^2({\cal A}^\prime, {\cal U}({\cal A}^\prime))}{d({\cal A}^\prime)}.
\end{align}
Namely, the $\cal A$-OTOC quantifies how far the commutant is mapped from itself by the dynamics. Clearly, this distance is maximal, i.e., maximally scrambling, when ${\cal A}^\prime$ gets orthogonalized to itself by the action of $\cal U$. 


We will now consider three important  physically motivated  examples, all of them collinear, in which the $\cal A$-OTOC can be fully computed analytically
(more details about the first two cases can be found in \cite{Zanardi-GAAC-2021}).

{\bf{1}}) {Let  ${\cal H}= {\cal H}_A\otimes {\cal H}_B$ be} a bipartite quantum system with ${\cal A}=L({\cal H}_A)\otimes \mathbf{1}_B$, and, therefore, 
 ${\cal A}^\prime=\mathbf{1}_A\otimes L({\cal H}_B).$ In this case one {recovers} the averaged bipartite OTOC discussed in \cite{styliaris_information_2020}
 \begin{align}\label{eq:op-ent}
 G_{\cal A}({\cal U})=1-\frac{1}{d^2}\langle  S_{AA^\prime},\,{\cal U}^{\otimes\,2}(S_{AA^\prime})\rangle,
 \end{align}
 where $d=d_A d_B=\mathrm{dim}\,{\cal H}\,\,(d_X=\mathrm{dim}\,{\cal H}_X\,(X=A,B))$, and $S_{AA^\prime}$ is the swap between the $A$ factors in ${\cal H}^{\otimes\,2}$.
 This a remarkable formula as it allows one to \emph{rigorously} prove a wealth of results  and unveil several relations between $G_{\cal A}(U)$ and other physical  quantities.

{\bf{2)}} Let ${\cal A}_B={\cal A}_B^\prime$ be the algebra of operators which are diagonal with respect to an orthonormal  basis $B:=\{|i\rangle\}_{i=1}^d$ i.e., ${\cal A}_B=\mathbf{C}\,\{ \Pi_i:=|i\rangle\langle i|\}_{i=1}^d.$
This is a $d$-dimensional maximal abelian subalgebra, and
\begin{align}\label{eq:MASA}
G_{{\cal A}_B}({\cal U})=1-\frac{1}{d}\sum_{i,j=1}^d|\langle i|U|j\rangle|^4.
\end{align}
This expression coincides with the \emph{coherence-generating power} (CGP) of $U$ introduced in  \cite{zanardiCoherencegeneratingPowerQuantum2017,styliaris_quantum_2019-1}. 
There, CGP is defined as the average coherence (measured
by the sum of squares of off-diagonal elements, with respect to $B$) generated by $U$ starting from any of the pure incoherent states $\Pi_i.$
Geometrically the CGP is  proportional to the distance  $D({\cal A}_B, {\cal U}({\cal A}_B))$ \cite{zanardiQuantumCoherenceGenerating2018} and 
has been applied to the detection of the localization transitions in many-body systems \cite{styliaris_quantum_2019-1}, and to detection of quantum chaos in closed  and open systems \cite{anand2020quantum}. 

{\bf{3)}} Let $\{S_l\}_{l=1}^{n-k}$ be a set of stabilizer operators over $n$-qubits \cite{stabilizer}, $G_S=\{ \prod_{l=1}^{n-k} S_l^{\alpha_l} \,|\, (\alpha_l)_l\in\{0,1\}^{n-k} \},$
the stabilizer group they generate and  ${\cal A}:=\mathbf{C}[G_S]$ its group algebra.
In Eq.~(\ref{eq:hilb-decomp}) now we have $J=(j_l)_l\in \{-1,\,1\}^{n-k},\,n_J=2^k,\,d_J=1.$  The VQS $\mathbf{C}^{n_J}$ with  $J=(1,\ldots,1)$ 
(identity irrep of $G_S$ and $\cal A$) is a {\em{quantum error correcting code}} \cite{stabilizer}. The $\cal A$-OTOC:
\begin{align}\label{eq:stab-A-OTOC}
G_{{\cal A}_S}({\cal U})=1-\frac{1}{2^{3n-k}}\sum_{g, h\in G_S}|\langle g, {\cal U}(h)\rangle|^2.
\end{align}
\subsection{Mereological Entropies}
It is important to stress  that $1-G_{\cal A}({\cal U})$ can be regarded as a {generalized purity \cite{gen-ent-2004}  and  that the $\cal A$-OTOC itself is therefore  a type of {{generalized linear entropy}}.

 For example, in the bipartite case {\bf{1)}} the $\cal A$-OTOC is {\em{identical}} to the operator linear entanglement entropy of $U$ as well as 
to  the average linear entropy production of the channel $T(\rho):= \mathrm{Tr}_B\left[{\cal U}(\rho\otimes\frac{\mathbf{1} }{d_B})\right]$ over pure $\rho$'s \cite{styliaris_information_2020}. This is an entropic contribution purely at the level of the {\em{multiplicative}} mereological structure. 

On the other hand, for abelian ${\cal A}^\prime={\cal Z}({\cal A})$,
the $\cal A$-OTOC  is an average  of the linear entropies of the probability vectors $\mathbf{p}_J:=\frac{1}{d_J}(\langle {\Pi_J}, {\cal U}(\Pi_K)\rangle)_{K=1}^{d_Z}$ 
\begin{align}\label{eq:entropic}
G_{\cal A}({\cal U})=\sum_J q_J S_L(\mathbf{p}_J),\quad q_J:={d_J}/{d},
\end{align}
where $S_L(\mathbf{p}_J):=1-\|\mathbf{p}_J\|^2,$ and $J=1,\ldots,d_Z.$
Since, from Eq.~(\ref{eq:A-OTOC-0}), one immediately sees that  $G_{\cal A}({\cal U})= G_{{\cal A}^\prime}({\cal U}^\dagger)$ for abelian $\cal A$, we have the same result with with $n_J$ and ${\cal U}^\dagger$ in lieu of $d_J$ and $\cal U$ respectively.
 These entropies measure how uniformly the $\Pi_K$'s are spread across the system of central sectors by the 
action of $U.$  This is an entropic contribution purely at the level of the {\em{additive}}  mereological structure.

For general $\cal A$, both the additive and the multiplicative mereological structures contribute to  the $A$-OTOC entropy. One has 
contributions from the ``uniformization'' of the the central sector system (due to the the "off-diagonal"  $U_{JK}:= \Pi_J U\Pi_K:$ for $J\neq K$) and the operator entanglement  of the individual $U_{JK}$ {(see \cref{app_entropies})}. 

\section{Scrambling  and subsystem emergence}\label{sec:rate}

We will now show how the $\cal A$-OTOC formalism developed so far can be used to define {{dynamical emergence of subsystems}}.  
Qualitatively, the idea is, given a Hamiltonian $H$ and an algebra $\cal A$, to define a notion  of ``scrambling time'' $\tau_s(H, {\cal A})$ (and ``scrambling rate'' $\tau_s^{-1}(H, {\cal A})$)   which is the time scale needed to have non-trivial scrambling, i.e, if $t\ll \tau_s$ no scrambling occurs. Once this is done, one can select algebras (gTPS)  by maximizing $\tau_s$ (minimizing $\tau^{-1}$) over a family $\mathbb{A}$ of operationally available $\cal A$'s. Symbolically
\begin{align}\label{eq:minimiz}
\{\textit{Emergent}\,{\cal A}\}:=
\mathrm{argmin}_{{\cal A}\in\mathbb{A}}\,\tau_s^{-1}(H, {\cal A})
\end{align}
Quantitatively, the starting point of our investigation is the following  result on the short-time dynamics of the $\cal A$-OTOC for Hamiltonian systems $U_t=e^{-i H t}.$

{\bf{Proposition}}{[\cref{appendix_proposition}]}
$G_{{\cal A}}({\cal U}_t)=2 (t/\tau_s)^2 + O(t^3)$, where the ``{{Gaussian scrambling rate}}"   $\tau_s^{-1} $ is given by
\begin{align}\label{eq:rate}
\tau^{-1}_s(H, {\cal A})= \|(1- \mathbb{P}_{{\cal A} +{\cal A}^\prime}) \tilde{H} \|_2 ,
\end{align}
where we have defined the normalized Hamiltonian $\tilde{H}:=\frac{H}{\sqrt{d}}.$
The ``gaussian scrambling time" $\tau_s$ 
can be regarded as defining  the period over which scrambling is negligible and  information ``locally" encoded in $\cal A$ is retained.
\par {The precise sense in which the Gaussian scrambling rate describes the rate of information scrambling with respect to both the additive and multiplicative structure of Eq. (\ref{eq:hilb-decomp}) becomes explicit by rewriting Eq. (\ref{eq:rate}) as (see \cref{appendix_eq15})}
 \begin{equation} \label{eq:split}
 \begin{split}
     &{\tau^{-2}_s(H,{\cal A})=}\\
     &{=\|(1-\mathbb{P}_{{\cal A}\vee{\cal A}^\prime})(\tilde{H})\|_2^2+  \| \mathbb{P}_{{\cal A}\vee{\cal A}^\prime}(1-\mathbb{P}_{{\cal A}+{\cal A}^\prime})(\tilde{H})\|_2^2.}
\end{split}
 \end{equation}
 The rate is given by two different types of contributions: coupling between different central sectors $\Pi_J{\cal H}\cong \mathbf{C}^{n_J}\otimes\mathbf{C}^{d_J}$ in (\ref{eq:hilb-decomp}), and  entangling ``interactions"  within them. {When $\mathcal{A}$ is a factor, the first contribution, which corresponds to the first term in Eq. (\ref{eq:split}), vanishes, and}
 \begin{equation} \label{eq:entang}
     \begin{split}
         &{\tau_s^{-1}=\left\lVert \tilde{H}+ \Tr(\tilde{H}) \frac{\mathbf{1}}{d} -\right.}\\
         &\hspace{40pt}{\left.-\frac{\mathbf{1}_{v_A}}{d_{v_A}} \otimes \Tr_{v_A}(\tilde{H})- \Tr_{v_B}(\tilde{H}) \otimes \frac{\mathbf{1}_{v_B}}{d_{v_B}} \right\rVert_2}
     \end{split}
 \end{equation}
 {where $\mathcal{H}\cong\mathcal{H}_{v_A} \otimes \mathcal{H}_{v_B}$ is the virtual bipartition induced by the dual pair of factors $(\mathcal{A},\mathcal{A}^\prime )$ and $d=d_{v_A} \, d_{v_B}$.}
 When either $\cal A$ or ${\cal A}^\prime$ is abelian, the second contribution{, which corresponds to the second term in Eq. (\ref{eq:split}}), vanishes, and one has
 $\tau^{-1}_s=\|(1-\mathbb{P}_{{\cal A}\vee{\cal A}^\prime})\tilde{H}\|_2,$ or even more explicitly
\begin{align}\label{eq:rate-abelian}
1/\tau_s^2={ \sum_J\| \Pi_J \tilde{H}(1-\Pi_J)\|_2^2}.
\end{align}
\par {Using Eq. (\ref{eq:rate}) we can derive a number of analytical properties of the gaussian scrambling rate:}
\par{\bf{Corollary}}{[\cref{appendix_cor}]} i) $G_{{\cal A}}({\cal U}_t)=0\,(\forall t)$ if (and only if)
$H$ belongs to the operator subspace ${\cal A} +{\cal A}^\prime:=\{a+b\,|\, a\in{\cal A},\,b\in{\cal A}^\prime\}.$ {Combined with Eq. (\ref{eq:rate}), this means that the $\mathcal{A}$-OTOC vanishes for all times if and only if $\tau_s^{-1}$ vanishes, despite the fact that the gaussian scrambling rate describes just the short-time expansion of $G_{{\cal A}}({\cal U}_t)$.}

ii) Using the standard Hilbert-space  notion of the distance between a vector $x$ and a subspace,
Eq.~(\ref{eq:rate}) can be recast into an even more geometrical fashion:
\begin{align}\label{eq:rate-distance}
\tau^{-1}_s(H, {\cal A})=D(\tilde{H},\,{\cal A} +{\cal A}^\prime).
\end{align}
 In words: for any   ${\cal A}$ the gaussian scrambling rate is  the  distance between the assigned (normalized) Hamiltonian and the sum
of the operator subspaces $\cal A$ and ${\cal A}^\prime.$ {Due to i),} when such a distance vanishes then the $\cal A$-OTOC is identically zero at {\em{all times}}.  

iii) The following upper-bounds hold:
\begin{align}\label{eq:tau-upper-bound}
\tau_s^{-1} (H, {\cal A})\le \mathrm{min}\{ D(\tilde{H}, {\cal A}),\,D(\tilde{H}, {\cal A}^\prime)\}\le \eta_H,
\end{align}
 where $\eta_H:=\|\tilde{H}\|_2$ is a natural energy scale associated with $H$  (whose inverse sets the physical unit of time).
 The first bound above implies  the nearly obvious  fact that if $H$ belongs to either $\cal A$ or ${\cal A}^\prime$ then there is no scrambling. Also, if either $\cal A$ or ${\cal A}^\prime$
 are abelian then the first upper-bound is saturated.
 On the other hand, Hamiltonians orthogonal  ${\cal A} +{\cal A}^\prime$ will saturate the last upper-bound and correspond to the {\em{fast scramblers}} in this short-time (gaussian) regime.
 
\vskip 0.2truecm
In the light of these results, the minimization problem (\ref{eq:minimiz}) has the following  neat (yet formal) geometrical solution:
{\em{the  dynamically ``emergent'' gTPS are those associated to the ${\cal A}\in \mathbb{A}$ such that the operator vector space ${\cal A}+{\cal A}^\prime$ has minimum distance from the Hamiltonian $H.$}} 
In particular, $\cal A$'s such that that distance is zero ($\eta_H$) correspond to no scrambling (fast scramblers). Moreover, when $\cal A$ (${\cal A}^\prime$) is abelian then ${\cal A}+{\cal A}^\prime$ in the above can be replaced by ${\cal A}^\prime$ ($\cal A$).

\subsection{Three Examples}
Let us now consider again the cases {\bf{1)}}-{\bf{3)}}  in the above. We will derive the explicit forms for $\tau_S(H, {\cal A})$ in these cases and show that in these important situations, Eq.~(\ref{eq:rate}) has a very physically transparent interpretation. This fact provides a key motivation to pursue the gaussian scrambling rate formalism in this paper.
%

%

{{\bf{1')}} {Spatial Bipartitions:}
Here we consider a state-space which is given a {background} multipartition into $N$  (ordered) qubits ${\cal H}=(\mathbf{C}^2)^{\otimes\,N}.$ If $S\subset \{1,\ldots,N\}$ (and $\bar{S}$ is its complement)  then, 
${\cal H} ={\cal H} _S\otimes {\cal H} _{\bar S}$, and the local observable algebras ${\cal A}_S:= L({\cal H} _S)\otimes\mathbf{1}_{\bar S}$ and  ${\cal A}_S^\prime= \mathbf{1}_{S}\otimes L({\cal H} _{\bar S}).$ Since $\mathbb{P}_{{\cal A}_S}(X)=\mathrm{Tr}_{\bar{S}} X\otimes \frac{\mathbf{1}}{d_{\bar{S}}},\,, \mathbb{P}_{{\cal A}_{\bar{S}}}(X)=\frac{\mathbf{1}}{d_{{S}}}\otimes\mathrm{Tr}_{{S}} X,$ and $\mathbb{P}_{{\cal A}_S} \mathbb{P}_{{\cal A}_{\bar{S}}}(X)=\frac{\mathrm{Tr}(X)}{d} \mathbf{1}$,
Eq.~(\ref{eq:rate}) takes the form:
\begin{align}\label{eq:rate-bipartite}
1/\tau_s =\| \tilde{H} -\frac{\mathbf{1}_S} {d_S}\otimes \mathrm{Tr}_S(\tilde{H}) -\mathrm{Tr}_{\bar{S}} (\tilde{H})\otimes \frac{\mathbf{1}_{\bar{S}}} {d_{\bar{S}}} \|_2,
\end{align}
where, for simplicity, we assumed a  traceless $H$ and $d_X=2^{|X|} ,(X=S, \bar{S}).$
Here we see that the projection $1-\mathbb{P}_{{\cal A}+{\cal A}^\prime}$ -- namely the projection onto $({\cal A}+{\cal A}^\prime)^\perp$ -- in Eq.~(\ref{eq:rate}) is nothing but the {{extraction of the $S$-${\bar S}$-interaction part of $H.$}}
This result leads to a particularly intuitive and satisfactory sense of ``subsystem emergence from minimal scrambling'':  {{given a family $\mathbb{S}$ of subsets of    $\{1,\ldots,N\}$, the Hamiltonian $H$ selects those  $S\in\mathbb{S}$  such that  $S$-${\bar S}$-interaction part of $H$ has minimal norm}}, i.e.,  those which are minimally coupled with the rest of the system.

{\bf{2')}} {Maximal Abelian Algebras:}
If ${\cal A}_B=\mathrm{span}\{|j\rangle\langle j|\} ={\cal A}_B^\prime,$ where the $|j\rangle$'s form a basis $B,$ then, from Eq.~(\ref{eq:rate-abelian}),
the gaussian scrambling rate   is the Hilbert-Schmidt norm of the {{off-diagonal}} (in the basis $B$) part  of $\tilde{H}.$ 
Intuitively, this quantity is as small as possible when the eigenbasis $B_H$ of $H$ and $B$ are as close as possible:
{{the dynamically selected ${\cal A}_B$'s are the closest to the one generated by $H.$}}
Indeed, 
if $H$ is non degenerate with an eigenbasis $B_H,$ one can prove  the bound 
\begin{align}\label{eq:A_B-bound}
\tau^{-1}_s(H, {\cal A}_B)\le\eta_H\, D({\cal A}_B,\,{\cal A}_{B_H}).
\end{align}
$D$  denotes the distance between algebras defined in the above and studied, in relation to CGP, in \cite{zanardiQuantumCoherenceGenerating2018}.
%

{\bf{3')}}  {{Stabilizer Algebra:}} 
$\mathbb{P}_{{\cal A}_S^\prime}(X)=|G_S|^{-1}  \sum_{g\in G_S} g X g^\dagger,$ and from {the Proposition} above,
\begin{align}\label{eq:rate-stab}
\tau^{-1}_s(H, {\cal A}_S)=\|(1 -\mathbb{P}_{{\cal A}_S^\prime}) \tilde{H}\|_2.
\end{align}
If $H$ anticommutes with any of the stabilizer generators $S_l$'s (see {\bf{3)}} above) then $\mathbb{P}_{{\cal A}_S^\prime}(\tilde{H})=0$ and the scrambling rate (\ref{eq:rate-stab}) is maximal.
This is exactly the condition for $H$ to generate correctable errors in the associate error correcting code. The elements
in ${\cal A}_S^\prime$ not in ${\cal A}_S$ correspond to non-trivial operations on the code or to uncorrectable errors \cite{stabilizer} and have $\tau^{-1}_s=0$.
%
%
\subsection{Rate minimization over families of algebras}\label{subsec:families-gen-TPS}
In this section we are going to illustrate the minimization strategy (\ref{eq:minimiz}) by means of  toy models  with ``circular'' families of algebras. In the first (second) case we have a {\em{continuous}} family of bipartitions (bases) of the two-qubit  (one-qubit) space. The third case corresponds to a family obtained by single-qubit rotations of two-qubit  symmetric  operators.
 
At the formal level, a  general way to build a family of algebras is to consider the adjoint orbit $\mathbb{A}:=\{{\cal A}_U := {\cal U}({\cal A}) \}_{U\in\mathbb{U}}$  of a given $\cal A,$ for $U$ ranging in some set  $\mathbb{U}$  of unitaries. Operationally, this means that the observer has the ability to implement elements of $\cal A$ {\em{and}} the unitaries in $\mathbb{U}$.
From (\ref{eq:rate-distance}) easily follows the covariance relation 
\begin{align}\label{eq:covariance}
\tau_s(H, {\cal U}({\cal A}))=\tau_s({\cal U}^\dagger (H), {\cal A}),
\end{align}
 which, in turn, shows that minimizing $\tau^{-1}_s(H,\bullet)$ over the adjoint orbit of $\cal A$ is equivalent to minimizing the scrambling rate of  $\cal A$ 
over an orbit of Hamiltonians $\mathbb{H}:=\{{\cal U}^\dagger (H) \}_{U\in\mathbb{U}}$, i.e., minimizing $\tau^{-1}_s(\bullet, {\cal A})\colon\mathbb{H}\rightarrow \mathbf{R}.$

{\bf{1'')}} Consider ${\cal H}=\mathbf{C}^2\otimes\mathbf{C}^2,$ with $H=\sigma^z\otimes\mathbf{1}$ and the family ${\cal A}_\theta:= U_\theta {\cal A}_0 U_\theta^\dagger,$ where
$ {\cal A}_0=L(\mathbf{C}^2)\otimes\mathbf{1},$ and $U_\theta:=e^{i\frac{\theta}{2}\sigma^x\otimes\sigma^x}.$
From Eq.~(\ref{eq:rate-bipartite})  $\tau^{-1}_s(H,  {\cal A}_{\theta})= |\sin\theta|.$
Non-trivial )vanishing rate for $\theta=\pi$ ($U_\pi\neq \mathbf{1}$ but ${\cal A}_0={\cal A}_\pi$) and gaussian fast scrambler for $\theta=\pi/2$ when the Hamiltonian is ``pure" interaction.

{\bf{2'')}}  Consider ${\cal H}=\mathbf{C}^2$, $H=\sigma^z$, and the  family ${\cal A}_{B_\theta}$, where $B_\theta=\{ U_\theta |0\rangle, \,U_\theta|1\rangle\}$ and $U_\theta:= e^{i\frac{\theta}{2}\sigma^y}.$
From v) and (\ref{eq:covariance}), one finds   $\tau^{-1}_s(H,  {\cal A}_{B_\theta})= |\sin\theta|.$ {This rate vanishes non-trivially} for $\theta=\pi$ ($U_\pi\neq \mathbf{1}$ but ${\cal A}_{B_0}={\cal A}_{B_\pi}$) and {is maximal, corresponding to a} gaussian fast scrambler, for $\theta=\pi/2$ when $B_{\pi/2}$ is mutually unbiased with respect to $B_0$ (and the rotated algebra ${\cal A}_{B_{\pi/2}}$ has maximal distance from ${\cal A}_{B_0}$).


{\bf{3'')}} Consider ${\cal H}=\mathbf{C}^2\otimes\mathbf{C}^2,$ with $H=\sigma^z\otimes\mathbf{1}+\mathbf{1}\otimes\sigma^z$, and the family ${\cal A}_\theta:= U_\theta {\cal A}_0 U_\theta^\dagger,$ where
$ {\cal A}_0=\mathbf{C}\{\mathbf{1},\,S\}^\prime$ ($S$ swap operator) and $U_\theta:=e^{i\frac{\theta}{2}\sigma^y}\otimes\mathbf{1}.$
From  v)  and (\ref{eq:covariance}),  one obtains  $\tau^{-1}_s(H,{\cal A}_\theta)=\frac{1}{4}\|H_\theta-SH_\theta S\|_2=\sqrt{2} \sin(\theta/2),$
where $H_\theta:=U_\theta^\dagger H U_\theta.$
Here  $\theta=\pi$  is a gaussian fast scrambler as the rotated Hamiltonian $H_{\pi}$ is now anti-symmetric and thus  orthogonal to ${\cal A}_0.$

While in these cases the optimization problem for $\tau_s^{-1}$ is straightforward, in general it might be a quite challenging one, even in physically intuitive cases.
Consider the case of minimizing $\tau_s^{-1}(H, \mathcal{A}_S)$ over spatial bipartitions of an $N$-qubit system, $\mathcal{H} = (\mathbb{C}^2)^{\otimes N}$, that may only contain \textit{two-qubit} interaction terms. We can represent $\mathcal{H}$ as a weighted graph $\mathcal{G} \equiv (V, E)$, with $V$ and $E$ being the set of vertices and edges respectively. Each of the $N = |V|$ vertices represents a qubit and each edge represents a two-qubit interaction term in the system Hamiltonian. We then have $\tilde{H} = 2^{-N/2}\sum_{(i,j)\in E}H_{ij}$.

In particular, let us consider the case of Ising interactions on each edge: $\tilde{H} = 2^{-N/2}\sum_{(i,j)\in E}J_{ij}\sigma_i^{\alpha}\otimes\sigma_j^{\beta}$, where $\alpha,\  \beta \in \{X,Y,Z\}$. For any subset of qubits $S \subset V$, the Gaussian scrambling rate, given by Eq. (\ref{eq:rate-bipartite}), is:
\begin{equation}
\tau_s^{-1}(H, \mathcal{A}_S) = \left(\sum_{(i,j)\in \partial S}J_{ij}^2\right)^{1/2},
\end{equation} where $\partial S \equiv \{(i,j)\in E | i\in S \text{ and } j \in \bar{S}\}$, the bipartition boundary. \footnote{Note that we arrive at this simple formula from the tracelessness of our interaction terms and the fact that they are factored into tensor products over our ``fundamental'' physical multipartition. The general expression for $\tau_s^{-1}(H, \mathcal{A}_S)$ is much more complicated for generic $h_{ij}$. For instance, for a 3-qubit system with boundary interaction  $\tilde{H} = J_{12}\ket{\phi^{+}}_{12}\bra{\phi^{+}} + J_{23}\ket{\phi^{+}}_{23}\bra{\phi^{+}}$, $\tau_s^{-1}(H, \mathcal{A}_S) = J_{12}^2+J_{23}^2 + J_{12}J_{23}/4$.}

\begin{figure*}
\centering
\begin{subfigure}{0.4\textwidth}
\centering
\includegraphics[scale=1.0]{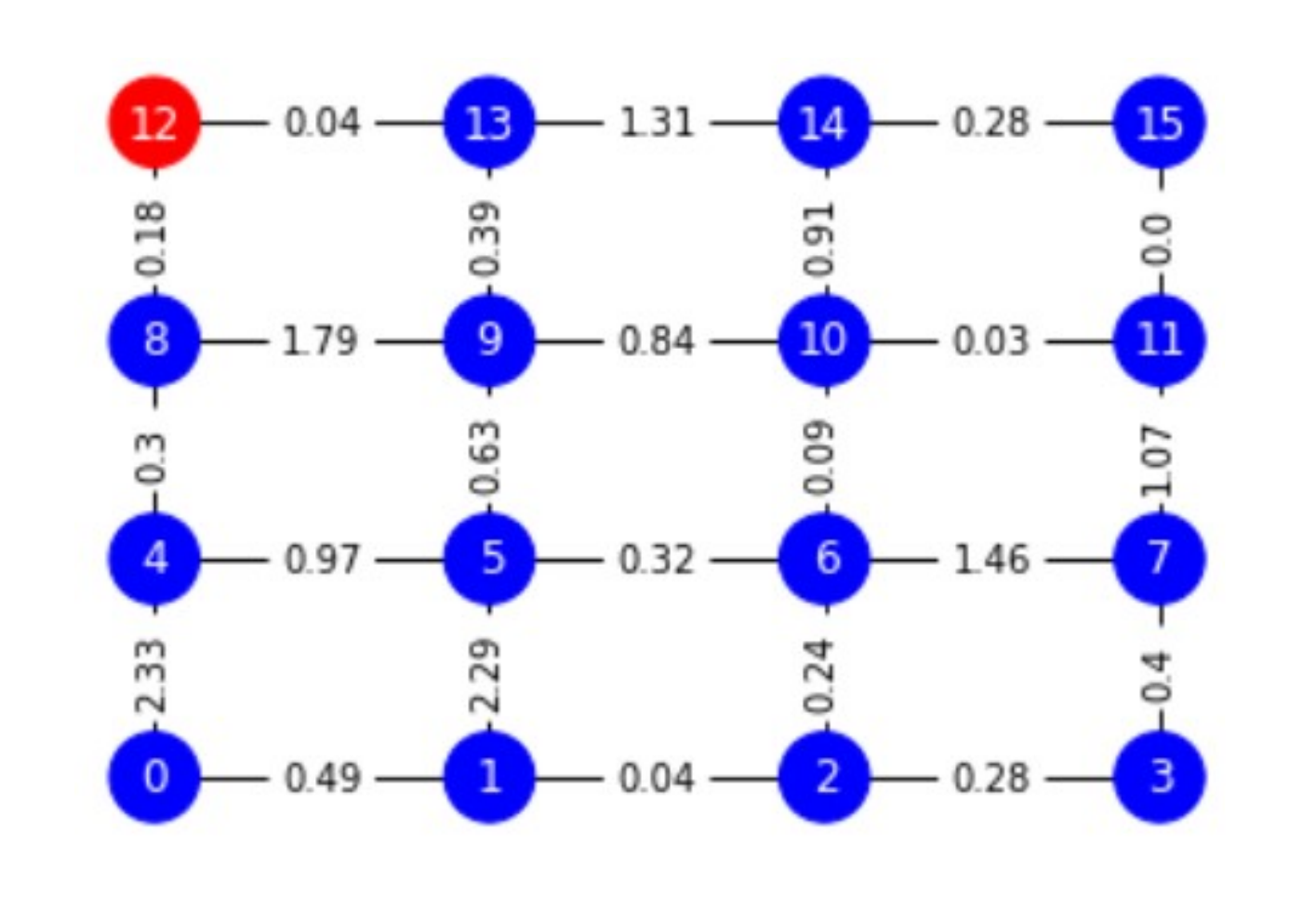}
\caption{}
\end{subfigure}\hspace{3pt}%
\begin{subfigure}{0.4\textwidth}
\centering
\includegraphics[scale=1.0]{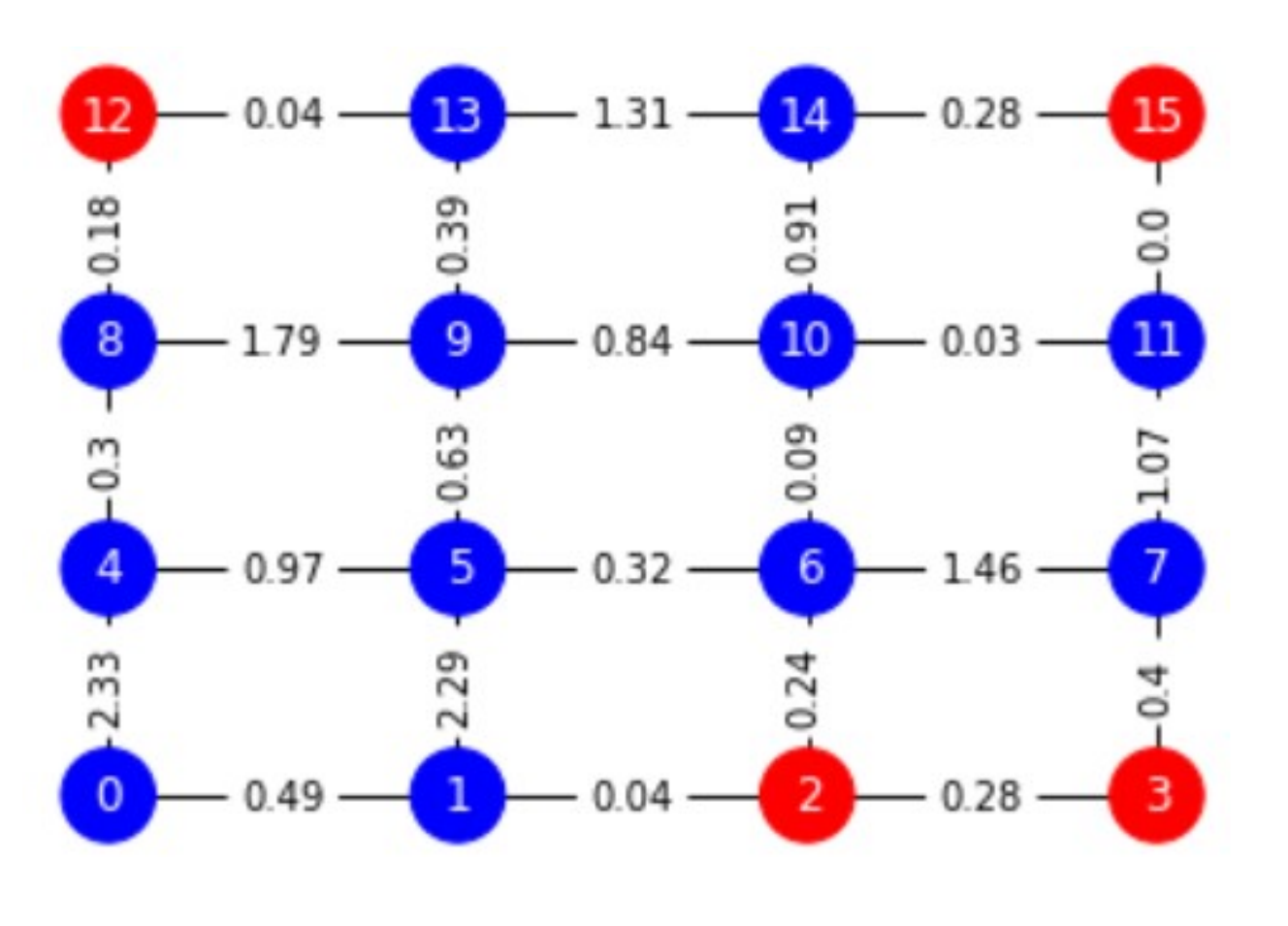}
\caption{}
\end{subfigure}
\caption{Figure a) shows the weighted min-cut bipartition of the system, with the subsystems $S$ and $\bar{S}$ denoted by red and blue nodes respectively. Figure b) shows the minimal $\tau^{-1}$ bipartition subject to the constraint that $|S| = 4$. Observe that the minimum cut in a) has a strictly lower weight than that of b) and can be obtained from b) by selecting just the lowest edge-weighted node in $S$ to be the new $S$.}
\label{fig_time}
\end{figure*}

The problem of minimizing the gaussian scrambling rate in this case corresponds precisely to solving the \textbf{weighted min-cut} problem in graph theory. This can be solved by the Stoer-Wagner algorithm, whose run-time complexity depends just on the number of vertices and number of edges \cite{stoer1997simple}.

In the special case of a graph with \textit{uniform} edge weights, we can normalize the Hamiltonian so that all $J$'s $= 1$. The graph is considered unweighted, with the corresponding problem known as \textbf{min-cut}. This problem is solved by the Karger-Stein algorithm \cite{10.1145/234533.234534, 10.5555/313559.313605}. Clearly, $\tau_s^{-1}(H, \mathcal{A}_S) = |\partial S|$. It is important to note that this unweighted minimization generally gives rise to a degenerate space of bipartitions. For example, in the case of a one-dimensional lattice, any bipartition consisting of a one-edged boundary yields $\tau_s^{-1}(H, \mathcal{A}_S) = 1$, and the solution space is $N-1$-fold degenerate.

To fully illustrate the problem, consider the special case of a two-dimensional qubit lattice, with edges from each qubit to its horizontal and vertical direct neighbors. Let us specify our interaction Hamiltonian as that of ZZ crosstalk, a common source of noise in superconducting quantum processors \cite{PhysRevLett.122.200502, 10.1145/3503222.3507761}: $\tilde{H} = \sum_{(i,j)\in E} J_{ij} \sigma_i^z\otimes\sigma_j^z$. For simplicity, we take the $J_{ij}$ to be normally distributed ($\mathcal{N}(0,1)$), and we consider at four-by-four grid of qubits. Figure \ref{fig_time} provides a graphical visualization.
\\

One may worry that the problem of minimizing the scrambling rate does not correspond exactly to \textbf{weighted min-cut} since the subsystems in our bipartition need not be simply connected. In the case that we minimize over \textit{all} possible bipartitions, it is obvious that any bipartition of disconnected subsystems can be decomposed into a bipartition of connected subsystems with a lower scrambling rate. On the other hand, if we constrain our family of bipartitions to those that contain equally sized subsystems, for example, the minimization becomes a more challenging task. Figure \ref{fig_time} demonstrates these distinct cases. Nonetheless, the direct mapping of the spatial bipartition scrambling rate minimization problem for a class of physically motivated Hamiltonians to a well-known graph theory problem demonstrates the feasibility of actually diagnosing the emergence of dynamically preferred subsystems.

We can probe properties of the minimum cut and resultant subsystems as parameters of our physical system are varied. Two quantities that easily admit further study are $\tau_S^{-1}$ itself, as well as typical subsystem size. We define $S$ to be the \textit{lower} cardinality subsystem in the min-cut solution.

It is perhaps natural to suspect that in the case of unconstrained $\tau_S^{-1}$ minimization, increasing the number of interacting qubit pairs as well as their interaction strengths will correspond one-to-one with smaller $|S|$; one may intuit that smaller $|S|$ implies a smaller $|\partial S|$, thereby decreasing the expected $\tau_S^{-1}$. However, we shall see this does not necessarily hold.

Again consider a lattice of qubits, now made boundary-less by adding interactions between opposite qubits on the original boundary, yielding a toroidal geometry. The $J_{ij}$ values between adjacent qubits are still drawn from $\mathcal{N}(0,1)$. Now, we introduce interactions between next nearest neighbors (NNN), which are one diagonal apart on the lattice, and assign to each a $J_{ij}$, also drawn from a normal distribution with a variable variance. We then increase the NNN interaction strengths (i.e. increase the variances of $J$) and see how average $|S|$ and $\tau_S^{-1}$ vary. Figure \ref{fig_2} contains plots of these relationships.

 To increase interactions, we generate NNN $J$ values by scalar multiplication of samples from $\mathcal{N}(0,1)$. It is thus expected that the $\tau_S^{-1}$ varies quadratically with the relative interaction strength of the NNN, since variance scales quadratically with scalar multiplication of a random variable, and edge weights are $J^2$ terms.

\begin{figure*}
\centering
\begin{subfigure}{0.3\textwidth}
\centering
\includegraphics[scale=1.0]{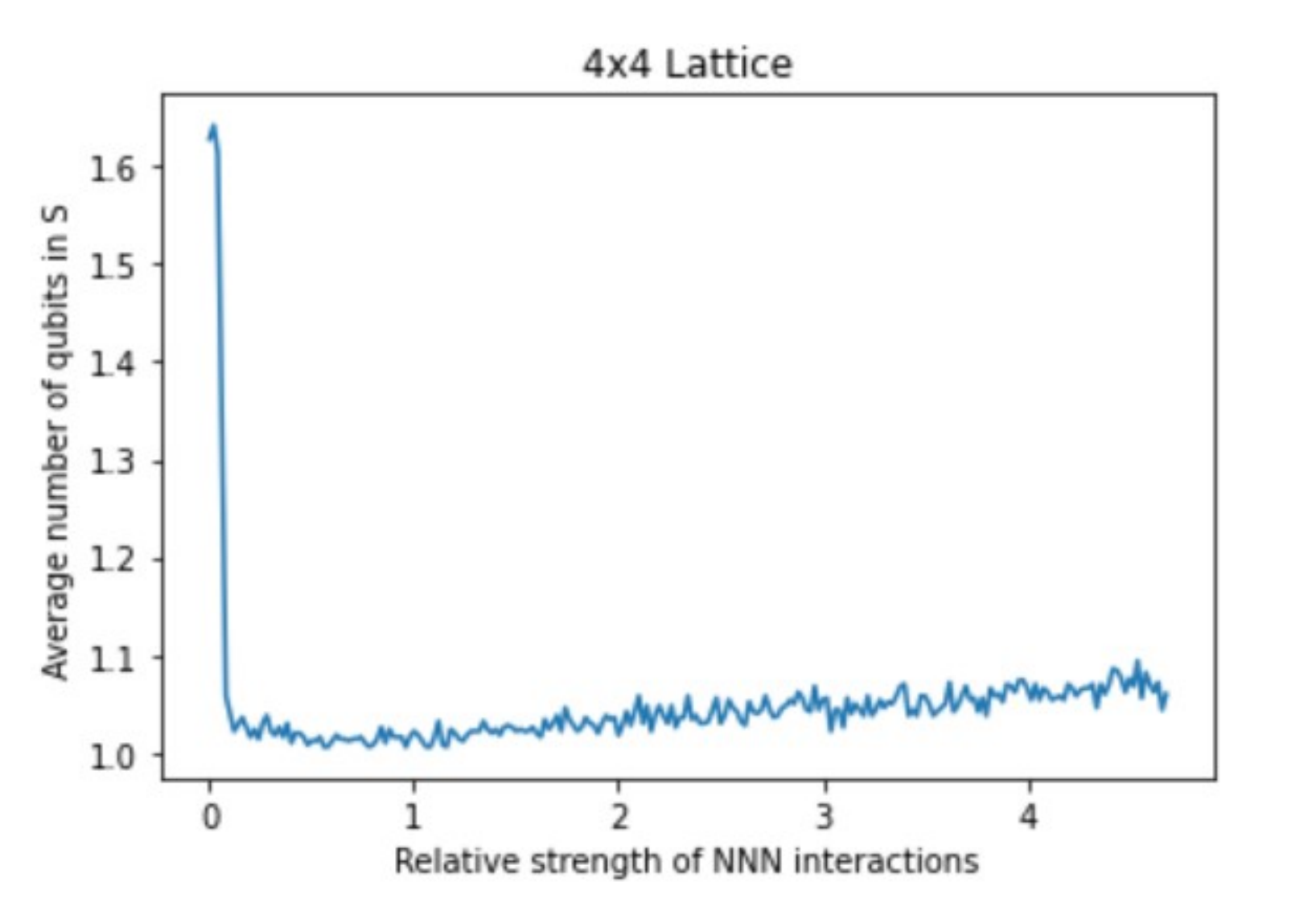}
\caption{}
\end{subfigure}\hspace{7pt}
\begin{subfigure}{0.3\textwidth}
\centering
\includegraphics[scale=1.05]{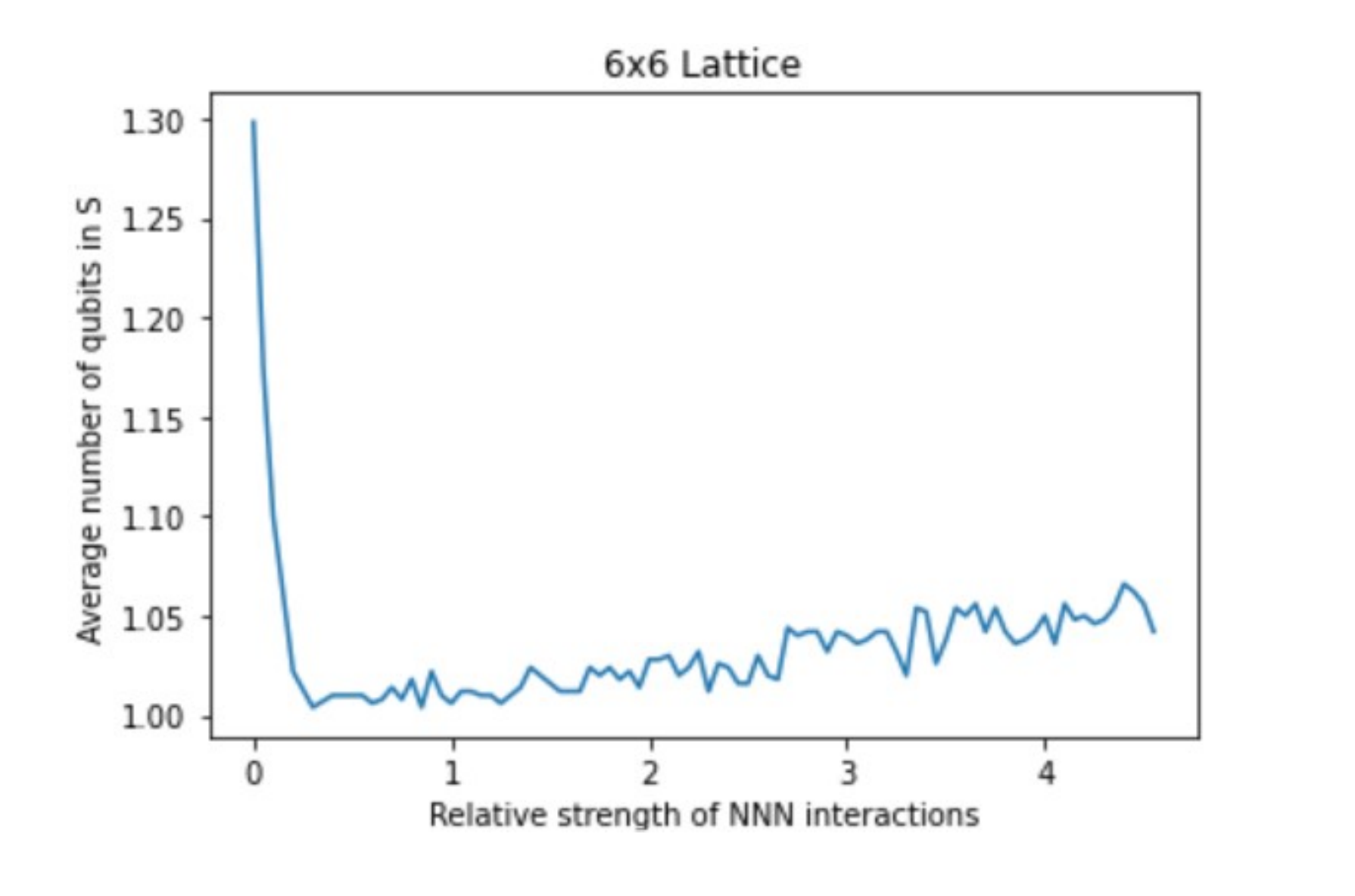}
\caption{}
\end{subfigure}\hspace{7pt}%

\caption{For each relative interaction strength $x$, 500 $J_{ij}$ samplings and corresponding runs of the min-cut algorithm were averaged. The relative interaction strength $x$ was implemented as a multiplier $x\cdot\mathcal{N}(0,1)$. X values were increased in increments of $0.02$, from $0$ to $5$, for the $4x4$ lattice, and in increments of $0.05$ for the $6x6$ lattice.}
\label{fig_2}
\end{figure*}

\begin{figure*}
\centering
\begin{subfigure}{0.3\textwidth}
\centering
\includegraphics[scale=1.0]{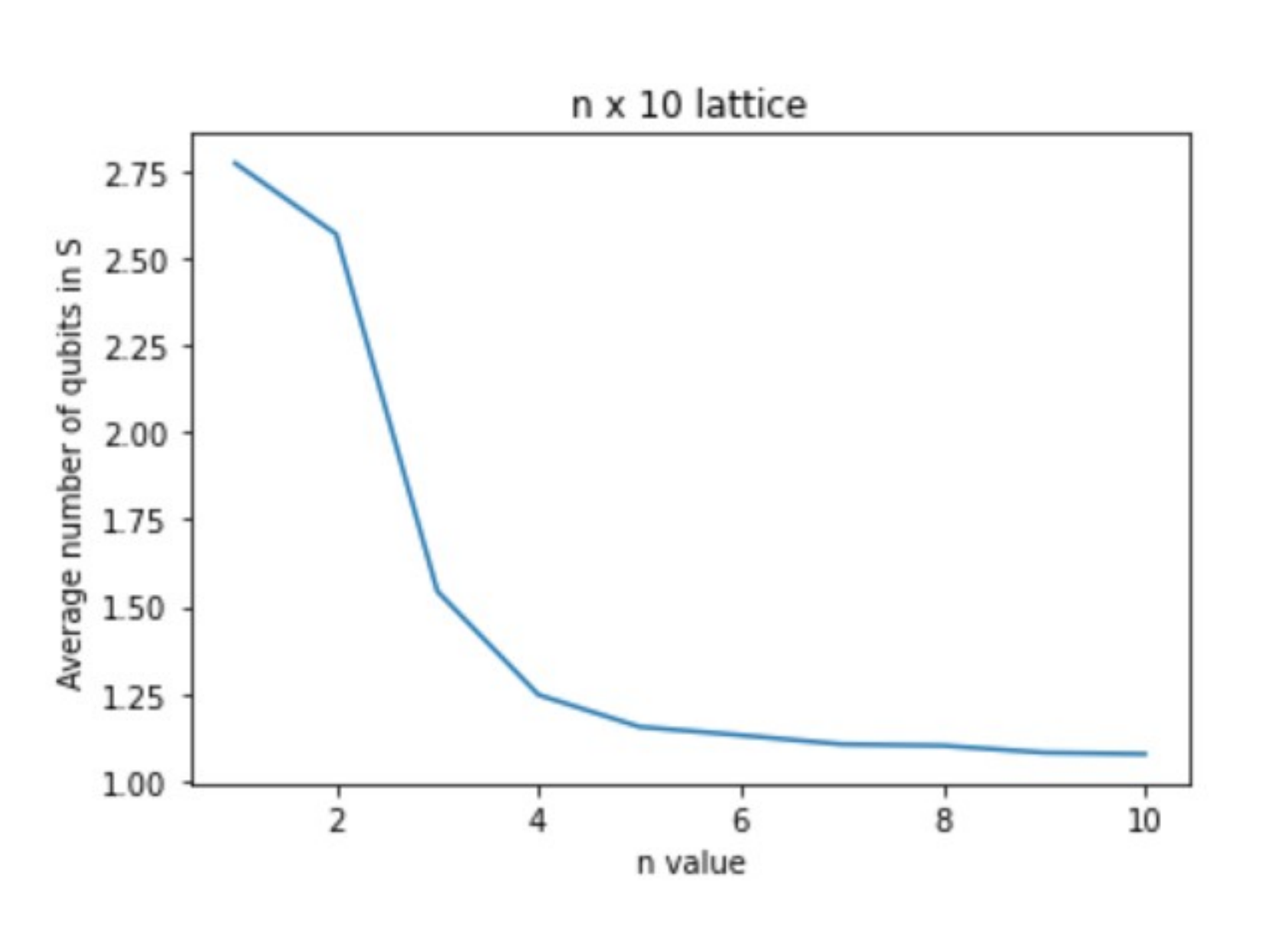}
\caption{}
\end{subfigure}\hspace{7pt}
\begin{subfigure}{0.3\textwidth}
\centering
\includegraphics[scale=1.0]{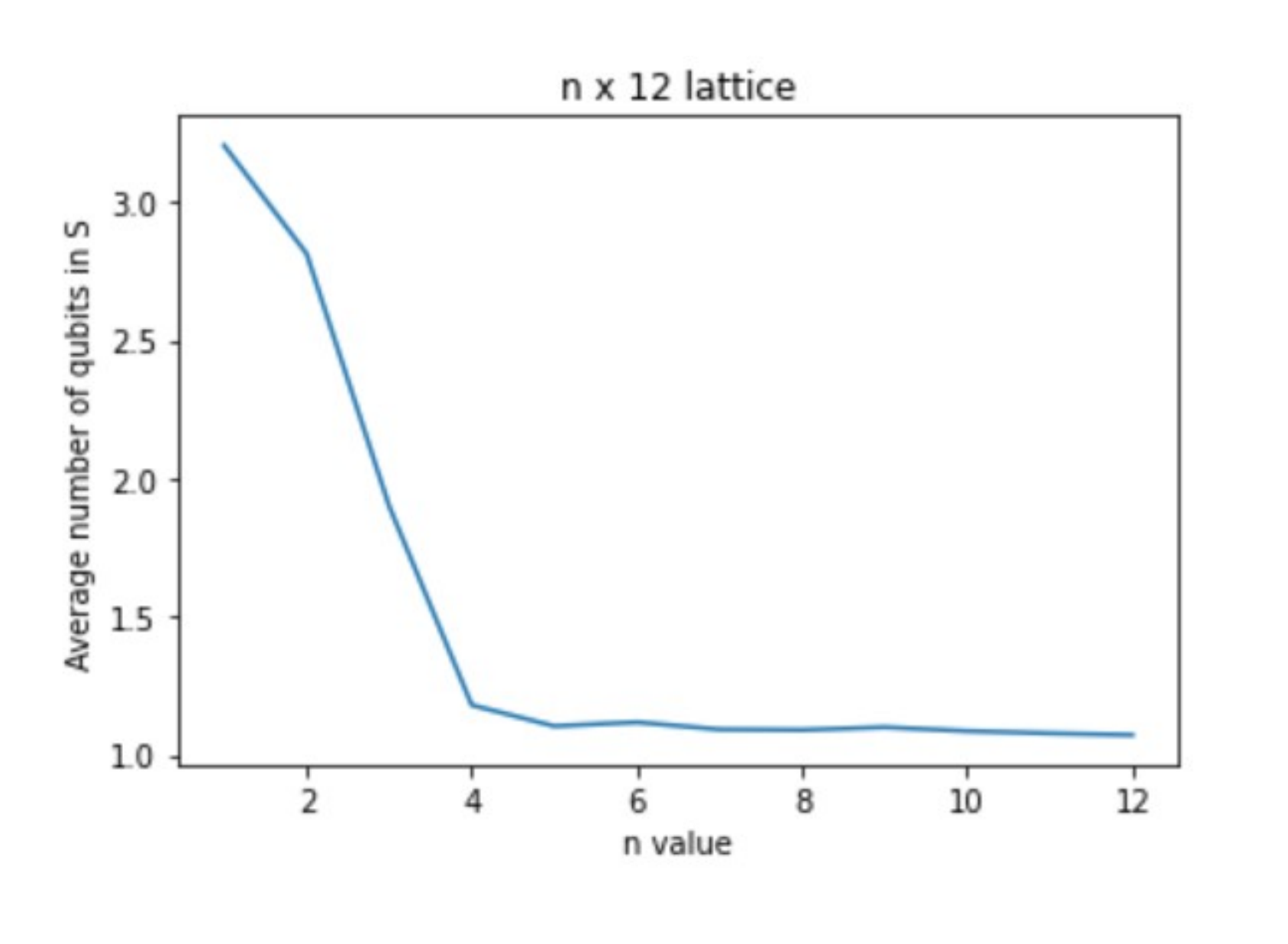}
\caption{}
\end{subfigure}\hspace{7pt}%

\caption{Again, 500 $J_{ij}$ samplings and corresponding runs of the min-cut algorithm were averaged. As n increases, the average size of $S$ approaches what we have seen previously.}
\label{fig_transition}
\end{figure*}
More interestingly, we see that after decreasing rapidly, $|S|$ begins to increase after the NNN interactions surpass the nearest-neighbor interactions in strength. When NNN interactions are very small, they contribute little to the total min-cut weight and can effectively be ignored. In the limit where NNN interactions are very strong, nearest-neighbor interactions contribute very little to the total min-cut weight and can effectively be ignored. In between, they both contribute significantly and the system is in a regime where $|\partial S| \propto \tau_S^{-1}$, thereby preferentially selecting smaller $S$ on average.

As the system size grows, the probability that there exists some particular qubit which has only weak interactions with its neighbors grows. It is thus expected that average $|S|$ decreases. This is observed in the difference between the $6x6$ and $4x4$ lattices in Figure \ref{fig_2}.

We can also probe how $|S|_{avg}$ changes as a lattice transitions from 1D to 2D. For a 1xN lattice, $|S|_{avg}$ should fall around $N/4$. We have seen that $|S|_{avg}$ falls between 1 and 2 for an NxN lattice. For n $\geq$ 3, we see  $|S|_{avg}$ rapidly approaches the NxN value, as the number of interactions per qubit is a constant 4 (\ref{fig_transition}).

\section{Conclusions}\label{sec:conclusions}
In this paper we have proposed a novel dynamical mechanism for {the} operational emergence of generalized tensor product structures.
 Generalized tensor product structures are defined by a dual pair $({\cal A},\,{\cal A}^\prime)$ made of a unital hermitian-closed subalgebra of implementable operators $\cal A$ and its commutant ${\cal A}^\prime$. This pair defines a Hilbert space mereology having both additive and tensor product components. When $\cal A$ is {a} factor, i.e., it has a trivial center, just the latter survives, and one has a standard virtual bipartition where the Hilbert space becomes a simple tensor product of two subsystems associated to $\cal A$ and ${\cal A}^\prime$ \cite{zanardi-virtual-2001}. 
 
 By means of a short-time expansion of the algebra OTOC recently introduced in \cite{Zanardi-GAAC-2021,A-OTOC-Faidon-2022}, {we} define a notion of gaussian scrambling time $\tau_s$, which is the time scale over which the system degrees of freedom associated with $\cal A$ are  (approximately) mapped onto themselves and no scrambling occurs. The rate $\tau_s^{-1}$ has a simple  and elegant geometric meaning in terms of the distance
 of the system Hamiltonian from the space ${\cal A}+{\cal A}^\prime$, and it involves couplings between different central sectors and intra-sector entangling interactions. 
 
 Given the system Hamiltonian, and a \textbf{family of operational possibilities}, our framework dynamically selects the {emergent} tensor product structures corresponding to maxima of $\tau_s$.
In the light of  the entropy-like nature of the algebra OTOC, this approach is reminiscent of the minimal entropy production principle and selects the structures which preserve their ``informational identity'' the longest.
\par {The intuition behind this idea becomes clear in some physically compelling examples we considered. Specifically, when $\mathcal{A}_B$ is a maximally abelian subalgebra corresponding to a basis $B$, the $\mathcal{A}$-OTOC quantifies the coherence-generating power of the unitary dynamics with respect to $B$ \cite{zanardiCoherencegeneratingPowerQuantum2017,styliaris_quantum_2019-1} and the gaussian-scrambling rate $\tau_S$ is closely related to the distance between the basis $B$ and the eigenbasis $B_H$ of a non-degenerate Hamiltonian $H$ that generates the time evolution. Given a family of operationally relevant bases (for example, a family of product bases in a quantum-many body setting), the minimization of $\tau_S$ provides a transparent notion of a dynamically preferred basis as the one that minimizes the rate with which coherence is generated. In addition, in the case that $\mathcal{A}$ is a factor, the $\mathcal{A}$-OTOC quantifies the operator entanglement of the unitary evolution operator $U$ \cite{styliaris_information_2020} across the bipartition induced by $\mathcal{A}$ and the gaussian scrambling rate corresponds exactly to the interaction strength between the subsystems of this bipartition. Minimizing the gaussian scrambling rate over operationally accessible bipartitions, then, describes the rather intuitive fact that the dynamics dictate a partitioning of the system into the subsystems that interact the weakest. We applied this framework in quantum many-body systems of qubits on a square lattice with random nearest neighbors (NN) and next nearest neighbors (NNN) Ising interactions. In this case, we mapped the minimization of $\tau_S^{-1}$ over bipartitions of the background qubits into a weighted min-cut problem of graph theory and studied the dependence of the average size of the emergent subsystems as we vary the relative strength of the NNN and NN interactions.}


{\section{Acknowledgments}}
FA, ED and PZ acknowledge partial support from the NSF award PHY-2310227. 
 This research was (partially) sponsored by the Army Research Office and was accomplished under Grant Number W911NF-20-1-0075.  SL was supported by DARPA under the RQMLS project, and by AFOSR and ARO.  The views and conclusions contained in this document are those of the authors and should not be interpreted as representing the official policies, either expressed or implied, of the Army Research Office or the U.S. Government. The U.S. Government is authorized to reproduce and distribute reprints for Government purposes notwithstanding any copyright notation herein.

\bibliographystyle{apsrev4-1}
\bibliography{refs, my_library}
\appendix
\begin{widetext}
\section{Commutativity and Information Transfer} \label{appendix_com}
Suppose  that for a unitary CP map ${\cal U}$ one has $[X, {\cal U}(Y)]=0,\,\forall X\in{\cal A},\forall Y\in{\cal B}$ (Heisenberg picture).

Let us prepare a family of states $\rho_\alpha=T_\alpha(\rho_0)$ where the $T_\alpha(\cdot):=\sum_i K_\alpha^i (\cdot)  (K_\alpha^i)^\dagger$ are trace preserving ($\sum_i (K_\alpha^i)^\dagger K_\alpha^i=\mathbf{1}$)  CP maps with Kraus operators in ${\cal A}.$ Now,  if $X\in{\cal B}$,  one has that (Schr\"odinger picture)
$\mathrm{Tr}\left[ X  {\cal U}^\dagger(\rho_\alpha)\right]=\mathrm{Tr}\left[ T_\alpha^\dagger {\cal U}(X)\rho_0\right]=\mathrm{Tr}\left[ {\cal U}(X)\rho_0\right].$
The last equality holds because,  since ${\cal U}(X)\in{\cal B}$ commutes with all operators in $\cal A,$ one has that $T_\alpha^\dagger {\cal U}(X)=\sum_i (K_\alpha^i)^\dagger   {\cal U}(X)K_\alpha^i
= (\sum_i (K_\alpha^i)^\dagger K_\alpha^i){\cal U}(X)={\cal U}(X).$  The above shows that the $\cal A$-local preparation maps $T_\alpha$ have no influence on the expectation values of operators in $\cal B.$ Namely,  the states ${\cal U}^\dagger(\rho_\alpha)$'s restricted to $\cal B$ are all identical and,  therefore,  no information can be sent by $\cal U$ from $\cal A$ to $\cal B$.

\section{The Factor Condition} \label{appendix_factor}
Let ${\cal A}\subset L({\cal H})$ be a unital hermitian-closed subalgebra. One has the surjective algebra homomorphism 
\begin{align}\label{eq:surj-homo}
{\cal A}\otimes{\cal A}^\prime\rightarrow {\cal A}\vee{\cal A}^\prime: a\otimes b\mapsto a b.
\end{align}
whose kernel is the ideal $\cal I$ of ${\cal A}\otimes{\cal A}^\prime$ generated by $\{ c\otimes \tilde{c}- \tilde{c}\otimes c\,|\, c,\tilde{c}\in{\cal Z}({\cal A})\}$ or, equivalently, by
$\mathrm{span}\{\Pi_J\otimes\Pi_K-  \Pi_K\otimes\Pi_J\}_{JK}$  where the $\Pi$'s are the central projections. Using standard isomorphisms one has that 
\begin{align}\label{eq:iso-quotient}
 \frac{ {\cal A}\otimes{\cal A}^\prime}{\cal I}\cong {\cal A}\vee{\cal A}^\prime.
\end{align}
Since $\mathrm{dim}({\cal I})=\sum_{J\neq K} d_J^2 n_K^2$ from (\ref{eq:iso-quotient}) one sees that the first isomorphism in Eq.~(\ref{eq:TPS-0}) is obtained if ${\cal I}=\{0\}$ i.e., there is just one $J$ in Eq.~(\ref{eq:hilb-decomp}). The latter condition means ${\cal Z}({\cal A})=\mathbf{C} 1$ which is the definition for $\cal A$ being a factor. In this case
$\mathrm{dim}({\cal A}\vee{\cal A}^\prime)=d_J^2 n_J^2= (n_J d_J)^2=\mathrm{dim}(L({\cal H}))$ which implies the second isomorphism in Eq.~(\ref{eq:TPS-0}).
\section{$A$-OTOC and entropies: the general case} \label{app_entropies}
i) In \cite{A-OTOC-Faidon-2022} in was proven that 
\begin{align}\label{eq:A-OTOC-1}
{G}_{{\cal A}}({\cal U}_t)=\frac{1}{d}\mathrm{Tr}\left[  S (1-\Omega_{\cal A}){\cal U}_t^{\otimes\,2}(\Omega_{{\cal A}^\prime})\right]=
1- \frac{1}{d}\mathrm{Tr}\left[  S \Omega_{\cal A}{\cal U}_t^{\otimes\,2}(\Omega_{{\cal A}^\prime})\right],
\end{align}
where $\Omega_{\cal A}:=\mathbb{E}_{X\in{\cal A}}[X\otimes X^\dagger]=\sum_{\alpha=1}^{d({\cal A})}e_\alpha\otimes e^\dagger_\alpha,\,$ $\Omega_{{\cal A}^\prime}:=\mathbb{E}_{Y\in{\cal A}^\prime}[Y\otimes Y^\dagger]=\sum_{\beta=1}^{d({\cal A}^\prime)}f_\beta\otimes f^\dagger_\beta,$ and $S$ is the swap operator
over ${\cal H}^{\otimes\,2}.$ 
The $e_\alpha$'s ($f_\beta$'s) form  an orthogonal (hermitean-closed) basis for $\cal A$ (${\cal A}^\prime$) and fulfill  \cite{Zanardi-GAAC-2021,A-OTOC-Faidon-2022}
\begin{align}\label{eq:Omega-Proj}
\mathrm{Tr}_1\left[ S\,\Omega_{\cal A}(X\otimes\mathbf{1}) \right]=\sum_\alpha e_\alpha X e_\alpha^\dagger=\mathbb{P}_{{\cal A}^\prime}(X),\quad
\mathrm{Tr}_1\left[ S\,\Omega_{{\cal A}^\prime}(X\otimes\mathbf{1}) \right]=\sum_\beta f_\beta X f_\beta^\dagger=\mathbb{P}_{{\cal A}}(X),
\end{align}
More specifically, $e_\alpha =\frac{ \mathbf{1}_{n_J}}{\sqrt{d_J}}\otimes |l\rangle\langle m|,\,\alpha:=(J, l, m);\, J=1, \ldots d_Z;\, l,m=1,\ldots, d_J$ and
$f_\beta= |p\rangle q|\otimes\frac{\mathbf{1}_{dJ}}{\sqrt{n_J}},\, p,q=1,\ldots, n_J$ from which it follows 
\begin{align}\label{eq:Omega-app}
\Omega_{\cal A}=\sum_\alpha e_\alpha\otimes e_\alpha^\dagger=\sum_J \mathbf{1}_{n_J}^{\otimes 2}\otimes \frac{S_{d_J}}{d_J},\quad
\Omega_{{\cal A}^\prime}=\sum_\beta f_\beta\otimes f^\dagger_\beta= \sum_J \frac{S_{n_J}}{n_J}\otimes  \mathbf{1}_{d_J}^{\otimes 2},
\end{align}
where $S_{d_J}\in L\left((\mathbf{C}^{n_J}\otimes  \mathbf{C}^{d_J})^{\otimes 2}\right)$ is the swap operators between the $ \mathbf{C}^{d_J}$ factors (similarly for the $S_{n_J}$'s.) Note that from (\ref{eq:Omega-app}) it follows $\sum_\alpha \|e_\alpha\|_2^2=\sum_\beta  \|f_\beta\|_2^2=d,$ and 
$\mathrm{Tr}[\Omega_{\cal A}]=\sum_J n_J^2=d({\cal A}^\prime),\, \mathrm{Tr}[\Omega_{{\cal A}^\prime}]=\sum_J d_J^2=d({\cal A}).$

Using Eq.~(\ref{eq:A-OTOC-1}), (\ref{eq:Omega-app}) and $S(\mathbf{1}_{n_J}^{\otimes 2}\otimes S_{d_J})=S_{n_J}\otimes \mathbf{1}_{d_J}^{\otimes 2},$ one gets the following expression for the $\cal A$-OTOC
\begin{align}\label{eq:general-entropic}
1-G_{{\cal A}}({\cal U})=\frac{1}{d}  \sum_{JK}\frac{1}{n_K d_J}\mathrm{Tr}\left[ ({S_{n_J}}\otimes  \mathbf{1}_{d_J}^{\otimes 2})\, {\cal U}_{JK}^{\otimes 2}( {S_{n_K}}\otimes  \mathbf{1}_{d_K}^{\otimes 2} )         \right],
\end{align}
where ${\cal U}_{JK}(X):= (\Pi_J U\Pi_K) X(\Pi_J U\Pi_K)^\dagger.$ For the factor case $d_Z=1$  one recovers the bipartite OTOC ( i.e., operator entanglement of $U$) (\ref{eq:op-ent}) while for abelian ${\cal A}^\prime$ ($n_K=1, \forall K)$ one finds: 
\begin{align}
1-G_{{\cal A}}({\cal U})=\frac{1}{d} \sum_{J}\frac{1}{ d_J}\sum_K\mathrm{Tr}\left[\Pi_J^{\otimes 2}{\cal  U}(\Pi_K)^{\otimes 2}\right]=
\sum_J q_J \sum_K |\langle\frac{\Pi_J}{d_J},\,{\cal U}(\Pi_K)\rangle|^2=\sum_J q_J \|\mathbf{p}_J\|^2
\end{align}
where $q_J:=\frac{d_J}{d_J}$ and $\mathbf{p}_J:=(\langle\frac{\Pi_J}{d_J},\,{\cal U}(\Pi_K)\rangle)_{K=1}^{d_Z}$ are probability vectors as $(\mathbf{p}_J)_K\ge 0$ and 
$\sum_K(\mathbf{p}_J)_K=\mathrm{Tr}[\frac{\Pi_J}{d_J}{\cal U}(\sum_K \Pi_K)]=\mathrm{Tr}[\frac{\Pi_J}{d_J}]=1.$ Then, using $\sum_Jq_J=\frac{1}{d}\sum_J d_J=1$ one obtains (\ref{eq:entropic}).

More in general, the term in (\ref{eq:general-entropic}) under trace can be seen to be proportional to the operator purity of the  maps $U_{JK}:= \Pi_J U\Pi_K:  {\cal H}_K\cong \mathbf{C}^{n_K}\otimes\mathbf{C}^{d_K} \rightarrow {\cal H}_J\cong \mathbf{C}^{n_J}\otimes\mathbf{C}^{d_J}.$ The entropic contributions to (\ref{eq:general-entropic})
are an intertwining of the ``uniformization" of the the central sector system (due to the the "off-diagonal"  $U_{JK}$ for $J\neq K$) and the operator entanglement  of the individual $U_{JK}.$ Both the additive and the multiplicative mereological structures contribute to  the $A$-OTOC entropic nature.
\section{Proof of the Proposition.}\label{appendix_proposition}
If ${\cal U}_t=e^{i t{\cal H}}$ by expanding the exponential one finds that the zeroth and first order terms vanish and 
$G_{{\cal A}}({\cal U}_t)=\frac{t^2}{d} \mathrm{Tr}\left[  (S (\Omega_{\cal A}-\mathbf{1}){\cal H}^{\otimes\,2}(\Omega_{{\cal A}^\prime})\right] +O(t^3).$ Now ${\cal H}(X)=HX-XH:=(L_H-R_H)(X)$ and therefore ${\cal H}^{\otimes 2}= L_H^{\otimes 2}- L_H\otimes R_H-R_H\otimes L_H+ R_H^{\otimes 2}.$
Computations using Eq.~(\ref{eq:Omega-Proj}) and $\mathrm{Tr}\left[ S(A\otimes B)\right]=\mathrm{Tr}[A B]$ lead to 
\begin{align}
\mathrm{Tr}\left[  S\, \Omega_{\cal A}{R_H}^{\otimes\,2}(\Omega_{{\cal A}^\prime})\right]=\mathrm{Tr}\left[  S\, \Omega_{\cal A}{L_H}^{\otimes\,2}(\Omega_{{\cal A}^\prime})\right]=\langle \mathbb{P}_{\cal A}(H), \mathbb{P}_{{\cal A}^\prime}(H)\rangle.
\end{align}
and
\begin{align}
\mathrm{Tr}\left[  S\, \Omega_{\cal A}({R_H}\otimes L_H)(\Omega_{{\cal A}^\prime})\right]= \mathrm{Tr}\left[  S\, \Omega_{\cal A}({L_H}\otimes R_H)(\Omega_{{\cal A}^\prime})\right]=
\| \mathbb{P}_{{\cal A}^\prime}(H)\|_2^2,\quad \mathrm{Tr}\left[ S {\cal H}^{\otimes 2}(\Omega_{{\cal A}^\prime})\right]=2\left(\|\mathbb{P}_{{\cal A}}(H)\|_2^2 -\|H\|_2^2\right).
\end{align}
Bringing all terms together (and up to higher order terms)
\begin{align}
G_{{\cal A}}({\cal U}_t)=\frac{2t^2}{d}\left( \|H\|_2^2-\|\mathbb{P}_{{\cal A}^\prime}(H)\|_2^2+   \langle \mathbb{P}_{\cal A}(H), \mathbb{P}_{{\cal A}^\prime}(H)\rangle -\|\mathbb{P}_{{\cal A}}(H)\|_2^2  \right) \rangle\nonumber \\=
\frac{2t^2}{d}\langle H, (1-\mathbb{P}_{{\cal A}})(1-\mathbb{P}_{{\cal A}^\prime})(H))\rangle
= \frac{2t^2}{d}\| (1-\mathbb{P}_{{\cal A}})(1-\mathbb{P}_{{\cal A}^\prime})(H)\|_2^2.
\end{align}
Now $\mathbb{P}_{{\cal A} +{\cal A}^\prime}= \mathbb{P}_{{\cal A}}+ \mathbb{P}_{{\cal A}^\prime}- \mathbb{P}_{{\cal A}}\mathbb{P}_{{\cal A}^\prime}$ and $1-\mathbb{P}_{{\cal A} +{\cal A}^\prime}=(1-\mathbb{P}_{{\cal A}})(1-\mathbb{P}_{{\cal A}^\prime})$ from which Eq.~(\ref{eq:rate}) follows.
\section{Proof of Eq. (\ref{eq:split})} \label{appendix_eq15}
Since ${\cal A}+{\cal A}^\prime\subset{\cal A}\vee{\cal A}^\prime,$  one has the identity $1-\mathbb{P}_{{\cal A}+{\cal A}^\prime}=(1-\mathbb{P}_{{\cal A}\vee{\cal A}^\prime}) + \mathbb{P}_{{\cal A}\vee{\cal A}^\prime}(1-\mathbb{P}_{{\cal A}+{\cal A}^\prime})$ one also finds 
\begin{align}\label{eq:rate-2-terms}
\tau^{-2}_s(H,{\cal A})=\|(1-\mathbb{P}_{{\cal A}\vee{\cal A}^\prime})(\tilde{H})\|_2^2+  \| \mathbb{P}_{{\cal A}\vee{\cal A}^\prime}(1-\mathbb{P}_{{\cal A}+{\cal A}^\prime})(\tilde{H})\|_2^2.
\end{align}
Since $  \mathbb{P}_{{\cal A}\vee{\cal A}^\prime}(X)=\sum_J\Pi_J X\Pi_J $ where the $\Pi_J$'s are the central projections in (\ref{eq:hilb-decomp}), the first term can be written $\sum_J\| \Pi_J H(1-\Pi_J)\|_2^2$ whereas  the second term is a sum over $J$ of terms as in Eq.~(\ref{eq:rate-bipartite}) squared.
The rate is given by two different types of contributions: coupling between different central sectors ${\cal H}_J\cong \mathbf{C}^n_J\otimes\mathbf{C}^{d_J}$ in (\ref{eq:hilb-decomp}) and  entangling ``interactions"  inside each of them. {When $(\mathcal{A},\mathcal{A}^\prime)$ is a dual pair of factors, the first contribution vanishes (the factor condition imposes that $\mathbb{P}_{\mathcal{A}\vee \mathcal{A}^\prime}=\mathbf{1}$) and one obtains (\ref{eq:entang})}. When either $\cal A$ or ${\cal A}^\prime$ is Abelian, the second contribution vanishes (${\cal A}\subset {\cal A}^\prime$ or ${\cal A}^\prime\subset {\cal A}$ implies  ${\cal A}+{\cal A}^\prime={\cal A}\vee{\cal A}^\prime$)
and one obtains (\ref{eq:rate-abelian}).
\section{Proof of the Corollary} \label{appendix_cor}
i) if $H\in {\cal A} +{\cal A}^\prime$ the ${\cal U}_t$ factorizes in two commuting unitaries, one acting on $\cal A$ and the other on ${\cal A}^\prime.$ Plugging this into the definition (\ref{eq:A-OTOC-0}) one sees that ${\cal U}(Y)\in{\cal A}^\prime$  and that the $\cal A$-OTOC then vanishes.
Viceversa, if the $\cal A$-OTOC is identically zero then also $\tau^{-1}_s=0$ which, by Eq.~(\ref{eq:rate}), is possible iff $H= \mathbb{P}_{{\cal A}+{\cal A}^\prime}(H)$ i.e., $H\in {\cal A} +{\cal A}^\prime.$ 
\vskip 0.2truecm
ii) Just use $D(x, W)= \inf _{y\in W} \|x-y\|=  \|(1-P_W)x\|.$ Where $P_W$ is the subspace projection.
\vskip 0.2truecm
iii) One has that $\|(1-\mathbb{P}_{{\cal A}})(1-\mathbb{P}_{{\cal A}^\prime})(H)\|_2\le \|(1-\mathbb{P}_{{\cal A}})(H)\|_2=D(H, {\cal A})$ and $\|(1-\mathbb{P}_{{\cal A}})(1-\mathbb{P}_{{\cal A}^\prime})(H)\|_2\le \|(1-\mathbb{P}_{{\cal A}^\prime})(H)\|_2=D(H, {\cal A}^\prime)$ from which the bound (\ref{eq:tau-upper-bound}) follows.
\section{Proof of Eq.~(\ref{eq:A_B-bound})}
The distance between two maximal abelian algebras ${\cal A}_B=\mathbb{C}\{|i\rangle\langle i|\}$ and ${\cal A}_{\tilde{B}}=\mathbb{C}\{|\tilde{i}\rangle\langle\tilde{ i}|\}$ is given
by $D({\cal A}_B, {\cal A}_{\tilde{B}})=\| \mathbb{P}_{{\cal A}_B}-   \mathbb{P}_{{\cal A}_{\tilde{B}}} \|_{HS}=\sqrt{ 2d(1-\frac{1}{d}\sum_{i,\tilde{i}} |\langle i|\tilde{i}\rangle|^4)}=
\sqrt{ 2d(1-\mathrm{tr}(X^T X)) }$ where $X_{i,j}:=|\langle i|\tilde{j}\rangle|^2$ \cite{zanardiQuantumCoherenceGenerating2018}.  From (\ref{eq:rate-abelian}) 
$$1/\tau_s^{2}(H, {\cal A}_B)=\frac{1}{d}\sum_{i=1}^d (\langle i|H^2|i\rangle -\langle i|H|i\rangle^2 )=\eta_H^2 (1- \langle \mathbf{\varepsilon}, X_H^T X_H \varepsilon\rangle), \quad \eta_H:=\frac{\|H\|_2}{\sqrt{d}}$$
where $H=\sum_k \tilde{\epsilon}_k |\tilde{k}\rangle\langle\tilde{k}|$ is the spectral resolution of $H,$ the normalized vector $\varepsilon=\frac{1}{\|H\|_2}(\tilde{\epsilon}_1,\ldots, \tilde{\epsilon}_d),$ and $(X_H)_{ik}:=|\langle i|\tilde{k}\rangle|^2.$ Then $\tau_s^{-2}\le \eta_H^2\| 1- X_H^T X_H\|_\infty\le \eta_H^2\,\| 1- X_H^T X_H\|_1=
 \eta_H^2 \,\mathrm{tr}\left( 1-  X_H^T X_H\right)=d\ \eta_H^2 \,\left(1-\frac{1}{d} \mathrm{tr}(X_H^T X_H)\right)=\frac{1}{2} \eta_H^2\, D^2( {\cal A}_B, {\cal A}_{B_H}),$ whence $\tau_s^{-1}\le {\eta_H}\,D( {\cal A}_B, {\cal A}_{B_H}) .$
\section{Proof of Eq.~(\ref{eq:covariance})}
By definition ${\cal U}({\cal A}):=\{ UxU^\dagger\,|\, x\in{\cal A}\}$ for unitary $U$ is a unital hermitian-closed subalgebra of $L({\cal H})$ if ${\cal A}$ is so.
Then, $\mathbb{P}_{{\cal U}({\cal A})}={\cal U}\mathbb{P}_{{\cal A}}{\cal U}^\dagger$ by direct inspection. The same relation is true for $({\cal U}({\cal A}))^\prime={\cal U}({\cal A}^\prime )$ i.e.,
$\mathbb{P}_{U{\cal A}^\prime U^\dagger}={\cal U}\mathbb{P}_{{\cal A}^\prime}{\cal U}^\dagger$
Using unitary invariance of $\|\cdot\|_2$ one finds:
\begin{align}
&&\tau_s^{-1}(H, {\cal U}({\cal A}))=\|(1-\mathbb{P}_{{\cal U}({\cal A}^\prime)})(1-\mathbb{P}_{{\cal U}({\cal A})})(H)  \|_2
=\| {\cal U} (1-\mathbb{P}_{{\cal A}^\prime }){\cal U}^\dagger {\cal U} (1-\mathbb{P}_{{\cal A}}) {\cal U}^\dagger(H) \|_2\nonumber \\&&=\|(1-\mathbb{P}_{{\cal A}^\prime })(1-\mathbb{P}_{{\cal A}}) {\cal U}^\dagger(H) \|_2
=\tau_s^{-1}({\cal U}^\dagger(H), {\cal A}).
\end{align}

\end{widetext}
\end{document}